\documentclass[aps,prb,amsmath,amssymb,reprint,superscriptaddress,preprintnumbers,showpacs,intlimits,longbibliography]{revtex4-1}
\usepackage{bm,latexsym,mathrsfs,enumerate,color}
\usepackage[mathcal]{euscript}
\usepackage[breaklinks=true,unicode=true,urlcolor = blue,colorlinks = true,citecolor = blue,linkcolor = blue]{hyperref}
\usepackage{graphicx}
\usepackage{todonotes}
\renewcommand{\vec}[1]{\bm{#1}}
\usepackage[normalem]{ulem}

\newcommand{\newtext}[1]{\textcolor[rgb]{0.00,0.00,0.00}{#1}}
\begin{document}
%
%
\title{Topologically stable magnetization states on a spherical shell: curvature stabilized skyrmions}

\author{Volodymyr P. Kravchuk}
\email{vkravchuk@bitp.kiev.ua}
\affiliation{Bogolyubov Institute for Theoretical Physics of National Academy of Sciences of Ukraine, 03680 Kyiv, Ukraine}
\affiliation{Leibniz-Institut f{\"u}r Festk{\"o}rper- und Werkstoffforschung, IFW Dresden, D-01171 Dresden, Germany}

\author{Ulrich K. R{\"o}{\ss}ler}
\affiliation{Leibniz-Institut f{\"u}r Festk{\"o}rper- und Werkstoffforschung, IFW Dresden, D-01171 Dresden, Germany}

\author{Oleksii M. Volkov}
\affiliation{Bogolyubov Institute for Theoretical Physics of National Academy of Sciences of Ukraine, 03680 Kyiv, Ukraine}

\author{Denis D. Sheka}
\affiliation{Taras Shevchenko National University of Kyiv, 01601 Kyiv, Ukraine}

\author{Jeroen~van~den~Brink}
\affiliation{Leibniz-Institut f{\"u}r Festk{\"o}rper- und Werkstoffforschung, IFW Dresden, D-01171 Dresden, Germany}

\author{Denys Makarov}
\affiliation{Helmholtz-Zentrum Dresden-Rossendorf e.V., Institute of Ion Beam Physics and Materials Research, 01328 Dresden, Germany}

\author{Hagen Fuchs}
\affiliation{Leibniz-Institut f{\"u}r Festk{\"o}rper- und Werkstoffforschung, IFW Dresden, D-01171 Dresden, Germany}

\author{Hans Fangohr}
\affiliation{University of Southampton, Southampton SO17 1BJ, United Kingdom.}

\author{Yuri Gaididei}
\affiliation{Bogolyubov Institute for Theoretical Physics of National Academy of Sciences of Ukraine, 03680 Kyiv, Ukraine}

\date{\today}

%
%
%
%
\begin{abstract}
Topologically stable structures include vortices in a wide variety of matter, such as skyrmions in ferro- and antiferromagnets, and hedgehog point defects in liquid crystals and ferromagnets. These are characterized by integer-valued topological quantum numbers. In this context, closed surfaces are a prominent subject of study as they form a link between fundamental mathematical theorems and real physical systems. Here we perform an analysis on the topology and stability of equilibrium magnetization states for a thin spherical shell with easy-axis anisotropy in normal directions. Skyrmion solutions are found for a range of parameters. These magnetic skyrmions on a spherical shell have two distinct differences compared to their planar counterpart: (i) they are topologically trivial, and (ii) can be stabilized by curvature effects, even when Dzyaloshinskii-Moriya interactions are absent. Due to its specific topological nature a skyrmion on a spherical shell can be simply induced by a uniform external magnetic field.
\end{abstract}

\maketitle

\section{Introduction}

Topological methods are increasingly used to describe observed states in condensed matter systems. Prominent examples are the description of vortex textures in superfluid helium;\cite{Anderson77,Volovik03} band theory for topological insulators;\cite{Hasan10,Moore10a,Hsieh08} topological
superconductivity in a helical Dirac gas\cite{Xu14} and in Dirac semimetals;\cite{Kobayashi15} and topological defects in liquid crystals,\cite{Alexander12,Kleman06}  ferromagnets,\cite{Thiele73,Belavin75,Malozemoff79,Papanicolaou91,Komineas96}and antiferromagnets.\cite{Barker16} In this context, thin curvilinear films of ordered matter are in the focus of strongly growing interest, because in these systems a nontrivial geometry can induce topological defects in the order parameter field\cite{Bowick09,Vitelli04,Turner10} and can result in new effective interactions.\cite{Napoli12,Napoli13,Gaididei14,Sheka15} Among curvilinear films the most promising candidates for new physical effects are closed surfaces due to the natural appearance of topological invariants in the system. In this case the normalized vector field $\vec{m}$ defined on the surface realizes a map of the surface into a sphere $S^2$. The degree $Q\in\mathbb{Z}$ of this map is an integer topological invariant,\cite{Mermin79,Thouless98,Dubrovin85p2} i.e.\ each given distribution of the vector field $\vec{m}$ on a closed oriented surface is characterized by an integer number $Q$ which is conserved for any continuous deformation (homotopy) of the field $\vec{m}$. Moreover, any two distributions of the field $\vec m$ are topologically equivalent (homotopic), i.e.\ they can be matched by means of a continuous deformation provided they have the same $Q$.\cite{Dubrovin85p2,Kosevich90,Manton04}
Since a discontinuity in the physical field $\vec{m}$ is usually energetically non-favorable, two solutions with different $Q$ are separated by a high energy barrier. This causes topological stability. For example, an isolated magnetic skyrmion\cite{Bogdanov89,Bogdanov94,Bogdanov99,Bogdanov01,Romming13,Romming15,Buettner15,Leonov16} in a planar film with Dzyaloshinskii-Moriya interaction (DMI) is an excited state of the system (for the case of low temperature and absence of external magnetic fields). However, this excitation is topologically stable, because the invariant is $Q=\pm1$ for the skyrmion,\cite{Papanicolaou91,Komineas15c} while $Q=0$ for the ground state.
Topological stability occurs for a variety of defects in ordered matter, such as disclination loops, hedgehog point defects and knots in nematic liquid crystals;\cite{Alexander12,Kleman06,Machon13,Jampani11,Senyuk12} and vortices\cite{Mertens00} and Bloch points\cite{Feldtkeller65b,Malozemoff79} in ferromagnets.

Conservation of the topological index $Q$ for a closed surface raises two fundamental questions: (i) what is the lowest energy equilibrium solution $\vec{m}(\vec{r})$ for a given $Q$, \newtext{which is not necessarily the ground state owing to the topological constraint.} And (ii) which $Q$ corresponds to the ground state for a given surface? The answers can lead to new phenomena, specific to the physical system under consideration. In this paper we answer these questions for the case of thin ferromagnetic spherical shells. Even such a relatively simple model brings a number of surprising results.

 We show that for a spherical shell a skyrmion solution exists as a topologically stable excitation above the hedgehog ground state. An important feature is that the skyrmion may be stabilized by curvature effects only, specifically by the curvature-induced, exchange-driven \emph{effective} DMI.\cite{Gaididei14,Sheka15} This is in contrast to the planar case, where the intrinsic DMI is required for the skyrmion stabilization.\cite{Bogdanov94,Sampaio13,Rohart13}

The case of the spherical shell is topologically opposite to that of the planar film: the skyrmion has the index $Q=0$, in other words it is topologically trivial, while the ground state  is characterized by  $Q=\pm1$. This is due to a shift of the topological index of the vector field, caused by topology of the surface itself. Since the skyrmion solution on a spherical shell is homotopic to a uniform state, it can be induced by means of a uniform external magnetic field, similarly to the excitation of onion magnetic states in nanorings.\cite{Rothman01} In a continuous medium the switching between states with different $Q$ is topologically forbidden. However, in discrete spin lattices such a transition is possible, though it requires a strong external influence.


\section{General case of an arbitrary curvilinear shell}
\begin{figure*}
	\includegraphics[width=\textwidth]{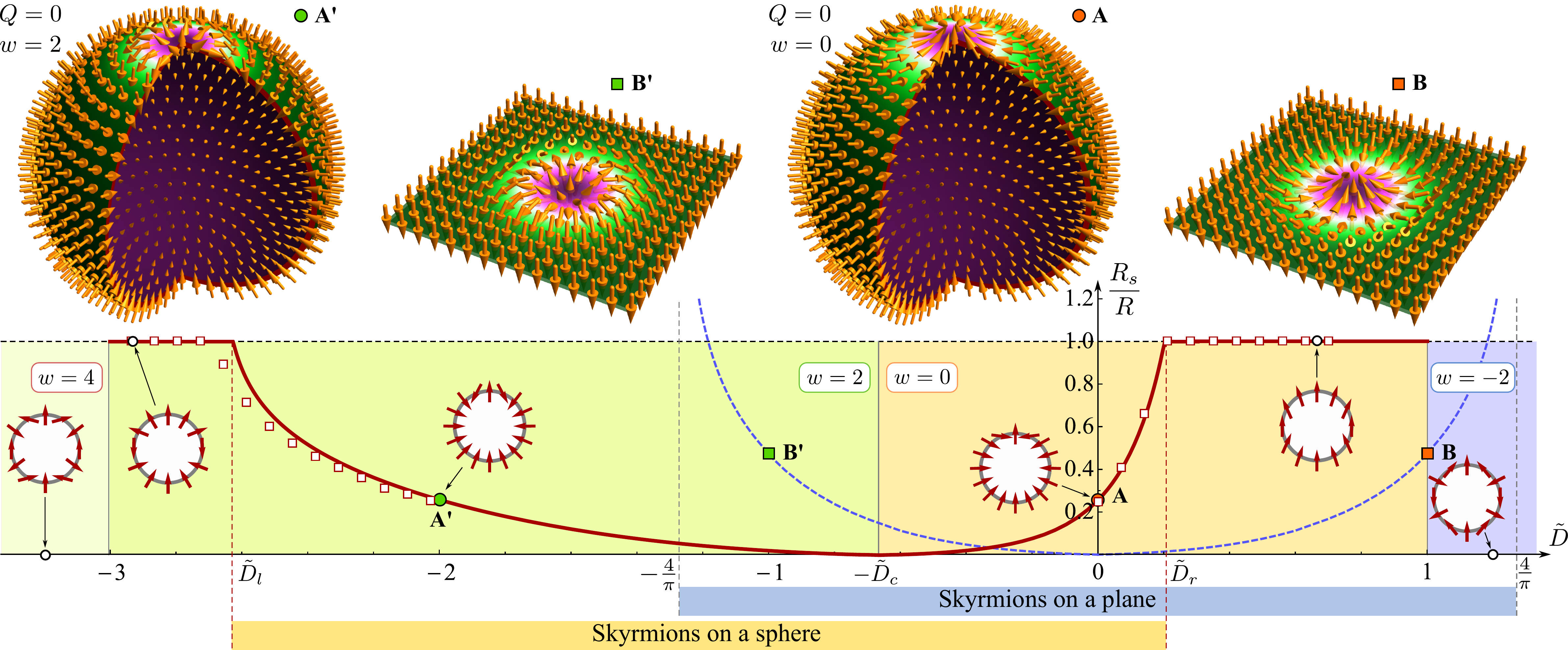}
	\caption{Topologically trivial magnetization states on a spherical shell \newtext{provide the answer to question (i) about the equilibrium (albeit not necessarily the ground) state with topological charge $Q=0$, as stated in the Introduction.} Insets A and A$'$ demonstrate the skyrmion solutions for different helicity numbers $w$. The respective skyrmion solutions for a planar film ($|Q|=1$) are shown in insets B and B$'$ for comparison. Case $A$ corresponds to the absence of the DMI ($D=0$). Dependences of the skyrmion radius $R_s$ on the dimensionless DMI constant $\tilde{D}=D/\sqrt{AK}$ for the case of a spherical shell (solid line) and for the case of a planar film (dashed line) are shown. \newtext{The skyrmion radii obtained from micromagnetic simulations are represented by open squares.} Small insets show the magnetization distribution along the vertical cross-section of the spherical shell. All calculations are performed for the case $R=3\ell$, where $\ell=\sqrt{A/K}$ is the characteristic magnetic length. The values of the DMI constant $\tilde D_l(R)$ and $\tilde D_r(R)$ determine the region of the skyrmion existence: $\tilde D\in(\tilde D_l,\,\tilde D_r)$.}\label{fig:Q0}
\end{figure*}

We first present a set of general results valid for an arbitrary thin curvilinear shell. In the following we apply these results to calculate the magnetic energy and topological properties of magnetization states of spherical shells.

\subsection{The mapping Jacobian} \label{sbsec:J}
The degree $Q$ of a map, realized by a normalized three-dimensional vector field $\vec{m}$ defined on a two-dimensional closed surface $\mathcal{S}$, reads\cite{Dubrovin85p2} $Q=(4\pi)^{-1}\int_{\mathcal{S}}\mathcal{J}\,\mathrm{d}\mathcal{S}$. In this particular case the mapping Jacobian $\mathcal{J}$ can be presented in the form of the triple product\cite{Dubrovin85p2}
$\mathcal{J}=-\epsilon_{\alpha\beta}\vec{m}\cdot\left[(\nabla_{\!\!\alpha}\vec{m})\times(\nabla_{\!\!\beta}\vec{m})\right]/2$, where the minus sign is introduced solely to conform with the traditional notation used in ferromagnetic research. Here and everywhere below the Greek indices $\alpha,\beta,...=1,2$ numerate the curvilinear coordinates $\xi_\alpha$, introduced on the surface, and the vector components defined in the corresponding curvilinear local basis $\vec{e}_\alpha$; while the Latin indices $i,j,k=1,2,3$ numerate coordinates and vector components in the Cartesian basis $\hat{\vec x}_i\in\{\hat{\vec x},\,\hat{\vec y},\,\hat{\vec z}\}$. The summation over repeated dummy indices is implied, unless stated otherwise. The local basis $\vec{e}_\alpha$ is assumed to be orthonormal $\vec{e}_\alpha\cdot\vec{e}_\beta=\delta_{\alpha\beta}$, therefore the metric tensor $||g_{\alpha\beta}||$ is diagonal. Details on the definition of the orthonormal basis for a given surface are presented in the Appendix~\ref{app:basis}. The operator $\nabla_{\!\!\alpha}\equiv (g_{\alpha\alpha})^{-1/2}\partial_\alpha$, where the summation over $\alpha$ is not implied and $\partial_\alpha=\partial/\partial\xi_\alpha$, denotes the corresponding component of the surface del operator
$\vec{\nabla}\equiv\vec{e}_\alpha\nabla_{\!\!\alpha}$. The surface element reads  $\mathrm{d}\mathcal{S}=\sqrt{g}\,\mathrm{d}\xi_1\mathrm{d}\xi_2$, where $g=\det||g_{\alpha\beta}||$.

Since using Cartesian components of the vector field $\vec{m}$ is not convenient for curvilinear systems, we will switch to curvilinear coordinates $\vec{m}=m_\alpha\vec{e}_\alpha+m_n\vec{n}$, where $\vec{n}=\vec{e}_1\times\vec{e}_2$ is the surface normal.
Moreover, it is useful to incorporate the constraint $|\vec{m}|=1$ by means of the angular parameterization $m_1+im_2 = \sin\theta e^{i\phi}$, $m_n=\cos\theta$, where $\theta=\theta(\xi_1,\xi_2)$ and $\phi=\phi(\xi_1,\xi_2)$ represent colatitude and longitude, the spherical angles of the local curvilinear basis, respectively.
In this case one can show (see Appendix~\ref{app:gyrovector}) that
\begin{equation}\label{eq:gyro-local}
\begin{split}
\vec{\mathcal{J}}\equiv \mathcal{J}\vec{n}=&-\sin\theta\,(\vec\nabla\theta-\vec\Gamma)\times(\vec\nabla\phi-\vec\Omega)-\\
&-\cos\theta\left[\left(\partial_\phi\vec\Gamma\times\vec\nabla\theta\right)+\vec n\mathcal{K}\right].
\end{split}
\end{equation}
Here, $\vec{\Gamma}(\phi)=||h_{\alpha\beta}||\cdot\vec{\varepsilon}(\phi)$, where $\vec{\varepsilon}=\cos\phi\vec{e}_1+\sin\phi\vec{e}_2$ is the normalized projection of the vector $\vec m$ on the tangential plane and $||h_{\alpha\beta}||$ is a tensor known as the Weingarten map or modified second fundamental form.\cite{Kamien02} Vector $\vec\Omega$ denotes the spin connection and $\mathcal{K}=\det||h_{\alpha\beta}||$ is the Gau{\ss} curvature. The corresponding definitions are presented in the Appendix~\ref{app:basis}.

One can easily check that for the case of a plane with a Cartesian frame of reference the expression \eqref{eq:gyro-local} results in the well known\cite{Thiele73,Belavin75,Malozemoff79,Papanicolaou91,Komineas96} formula $\vec{\mathcal{J}}=\vec\nabla(\cos\theta)\times\vec\nabla\phi$.

Remarkably, for a strictly normal distribution of the vector field $\vec{m}=\pm\vec{n}$ (normal Gau{\ss} map) one obtains the well known\cite{Kamien02,Dubrovin85p2} result $\mathcal{J}=\mp\mathcal{K}$.  Applying the Gau{\ss}-Bonnet theorem we obtain the famous relation $Q_\textsc{g}=\mp(1-\mathfrak{g})$ between degree of the normal Gau{\ss} map $Q_\textsc{g}$ and genus $\mathfrak{g}$ of the surface. Thus, $Q_\textsc{g}=\pm1$ for a normally magnetized sphere (hedgehog), $Q_\textsc{g}=0$ for a normally magnetized torus, etc. 
\newtext{In a topological classification of the solutions the value $Q_\textsc{g}$ should be taken into account as a topological charge shift, which originates from the topology of the surface itself.  To establish a link with the well-known skyrmions in the planar geometry\cite{Bogdanov89,Bogdanov94,Bogdanov99,Bogdanov01,Romming13,Romming15,Buettner15,Leonov16} one has to introduce the skyrmion number $\mathcal{N}\equiv Q-Q_\textsc{g}$.\footnote{For a planar film $Q_\textsc{g}=0$.} In the following, we consider skyrmions with $\mathcal{N}=\pm1$.  However, in the general case $\mathcal{N}$ can be an arbitrary integer not equal to zero.}

\newtext{Note that the term ``skyrmion'' is used rather broadly: any localized two-dimensional structure with unit (integer) mapping degree $Q$ may be considered a skyrmion. However, in addition to chiral skyrmions\cite{Bogdanov89,Bogdanov94,Bogdanov99,Bogdanov01,Romming13,Romming15,Buettner15,Leonov16} and bubbles,\cite{Malozemoff79} this definition includes a variety of objects with very different physical properties, such as vortex domain walls on tubes;\cite{Landeros10,Yan11a,Villain-Guillot95} hedgehog states and some vortex states on a spherical shell;\cite{Milagre07,Kravchuk12a} and rotating vortex dipoles.\cite{Komineas07a}
It is instructive to introduce a narrower definition which considers skyrmions as localized solutions with the structure of a vortex.\footnote{In fact, seminal works on magnetic skyrmions,\cite{Bogdanov89,Bogdanov94,Bogdanov99,Bogdanov01} used the term ``vortex'' instead of ``skyrmion''.}}

The vector $\vec{\mathcal{J}}$ is the limit for the two-dimensional case for the gyrocoupling vector\cite{Thiele73,Malozemoff79,Papanicolaou91,Komineas96}(topological density, topological current, vorticity) $\vec{\mathrm{J}}$, whose Cartesian components read $\mathrm{J}_i=-\epsilon_{ijk}\vec{m}\cdot[\partial_j\vec m\times\partial_k\vec m]/2$. The gyrocoupling vector is widely used for the topological description of a unit vector field $\vec{m}$ defined in a three-dimensional domain. If the shell thickness $L\to0$ is small enough to ensure the uniformity of $\vec{m}$ along the normal direction: $\vec{m}=\vec{m}(\xi_1,\xi_2)$, then $\vec{\mathrm{J}}\to\vec{\mathcal{J}}$, see Appendix~\ref{app:gyrovector}.
In magnetism, the gyrocoupling vector $\vec{\mathrm{J}}$ is the key quantity to describe the dynamics of topologically nontrivial solutions, such as domain walls,\cite{Thiele73,Malozemoff79} vortices,\cite{Huber82a,Papanicolaou91,Mertens00} skyrmions,\cite{Komineas15c,Lin13,Lin13b,Lin15a} skyrmion lines\cite{Milde13,Lin16} and Bloch points.\cite{Malozemoff79,Pylypovskyi12,Milde13,Lin16} It determines important integrals of motion in the dynamics of ferromagnetic media.\cite{Papanicolaou91,Komineas96} Recently it was shown\cite{Schulz12,Lin16} that $\vec{\mathrm{J}}$ is proportional to the emergent magnetic field, which appears due to the Hund's coupling between spins of the conducting electrons and localized magnetic moments. This gives rise to the topological Hall effect.\cite{Neubauer09,Li13a,Kanazawa11}

Let us provide physically illustrative explanations why the topological charge or index $Q$ is an integer number and a conserved quantity. A direct consequence of the definition of $\vec{\mathrm{J}}$ with the constraint $|\vec{m}|=1$ is $\mathrm{div}\,\vec{\mathrm{J}}=-4\pi\sum_nc_n\delta(\vec r-\vec R_n)$,
where $\delta(\vec r)$ is the Dirac delta-function and the vector $\vec R_n$ determines the position of a Bloch point (monopole), $\vec{m}_n^B$, whose infinitesimal neighborhood of the center has the structure\cite{Malozemoff79}
$\vec{m}_n^B=c_n\mathfrak{R}(\vec r-\vec R_n)/|\vec r-\vec R_n|$.
Here, $\mathfrak{R}$ is an arbitrary matrix of three-dimensional rotations and $c_n=\pm1$ is the monopole charge. Thus, the monopoles are sources and sinks of the gyrovector field.\cite{Malozemoff79}  Likewise, electrical charges are sources and sinks of the electrical field.
For any closed surface $\mathcal{S}$ enclosing the volume $\mathcal{V}$, the integral
$Q=(4\pi)^{-1}\int_{\mathcal{S}}\vec{\mathrm{J}}\cdot\vec{\mathrm{d}\mathcal{S}}=(4\pi)^{-1}\int_{\mathcal{V}}\mathrm{div}\,\vec{\mathrm{J}}\,\mathrm{d}\mathcal{V}=-\sum_nc_n$
yields an integer number $Q\in\mathbb{Z}$ equal to the difference of negatively and positively charged monopoles inside $\mathcal{S}$. Since two monopoles with opposite charges are connected by the Dirac string, which may be considered a skyrmion line,\cite{Milde13,Lin16} one can also say that $Q$ is the difference of outgoing and incoming skyrmion lines.\cite{Milde13} Thus the only way to change $Q$ for a given closed $\mathcal{S}$ is to replace a monopole across $\mathcal{S}$. When the monopole center crosses the surface, i.e.\ it is located exactly on $\mathcal{S}$, the vector field $\vec m_{\mathcal{S}}(\vec{r})=\vec{m}(\vec{r}\in\mathcal{S})$ is discontinuous at the point $\vec R_n$. Thus, one can conclude that \emph{a continuous deformation of the continuous field $\vec m_{\mathcal{S}}(\vec{r})$ can not change the mapping degree $Q$ of $\mathcal{S}$}. The rigorous proof of the latter statement can be found elsewhere, for instance see Ref.~\onlinecite{Dubrovin85p2}.

The expression for the mapping Jacobian \eqref{eq:gyro-local} is general: it is the key formula for topological analysis of a normalized vector field of an arbitrary physical nature on an arbitrary curvilinear surface.

\subsection{Magnetic energy of a curvilinear shell}\label{sbs:model}

 The topological analysis is independent of the physical nature of the vector field $\vec{m}$. In the following, we focus on possible equilibrium magnetization states of thin ferromagnetic curvilinear shells. To this end, we introduce the energy functional $E=L\int\left[A\,\mathscr{E}_{\mathrm{ex}}-K(\vec{m}\cdot\vec{n})^2+D\,\mathscr{E}_{\textsc{d}}\right]\mathrm{d}\mathcal{S}$.
Here, we take into account three magnetic interactions. The first term of the integrand represents the exchange energy with the energy density $\mathscr{E}_{\mathrm{ex}}=\partial_i\vec{m}\cdot\partial_i\vec{m}$ and the exchange constant, $A$. The second term is a uniaxial anisotropy: easy-normal for $K>0$ or easy-surface for $K<0$. \newtext{The presence of this anisotropy, which conforms to the geometry, is crucial for our model.  The anisotropy forces spins to follow the geometry which is why the spin subsystem ultimately ``fills'' the geometry.  This is a fundamental difference between our approach and a number of previous studies, where soliton solutions were found on curvilinear shells, yet anisotropy was either neglected,\cite{Villain-Guillot95,Carvalho-Santos12,priscila15,Carvalho-Santos15a} or it was spatially uniform lacking any correlation with the geometry.\cite{Milagre07,Carvalho-Santos08,Carvalho-Santos13} Our approach is based on the fundamental behavior of magnetically ordered media, where spin-orbit couplings provide the vital link between nontrivial curved geometry and the spin-system. Therefore, any realistic assessment of possible magnetization states in curved geometries must include the geometrically allowed anisotropic couplings.}

The last term in $E$ is the DMI with the energy density\cite{Crepieux98,Bogdanov01,Thiaville12} $\mathscr{E}_{\textsc{d}}=m_n\vec{\nabla}\cdot\vec{m}-(\vec m\cdot\vec{\nabla})m_n$ and the DMI constant, $D$. This kind of DMI originates from the spin-orbit coupling and is related to the inversion symmetry breaking on the film interface; it is typical for ultrathin films\cite{Crepieux98,Bogdanov01,Thiaville12} or bilayers.\cite{Yang15} 
In the curvilinear basis one can represent the DMI density as follows
\begin{equation}\label{eq:Ed}
\mathscr{E}_{\textsc{d}}=\vec{\varepsilon}\cdot\vec{\nabla}\theta+\sin\theta\cos\theta\,\partial_\phi\vec{\varepsilon}\cdot\left(\vec{\nabla}\phi-\vec{\Omega}\right)-\mathcal{H}\cos^2\theta,
\end{equation}
where $\mathcal{H}=\mathrm{tr}||h_{\alpha\beta}||$ is the mean curvature, see Appendix~\ref{app:interactions} for details. It is clear from \eqref{eq:Ed}, that an effective uniaxial anisotropy along $\vec{n}$ appears with a coefficient equal to $\mathcal{H}$.

\newtext{In our model we assume that the magnetostatic interaction, which is always present in the system, can be reduced to the easy-surface anisotropy, resulting in the shift of the anisotropy coefficient $K$. This was rigorously demonstrated\cite{Gioia97,Kohn05,Kohn05a}  
for plane films, when  thickness $L$ is substantially smaller than the size of the system and $L\ll\sqrt{A/(4\pi M_s^2)}$. Here, we assume that the same model is sufficient for smoothly curved shells, if $L$ is much smaller than the curvature radius.\cite{Slastikov05}}

\begin{figure*}
	\includegraphics[width=0.9\textwidth]{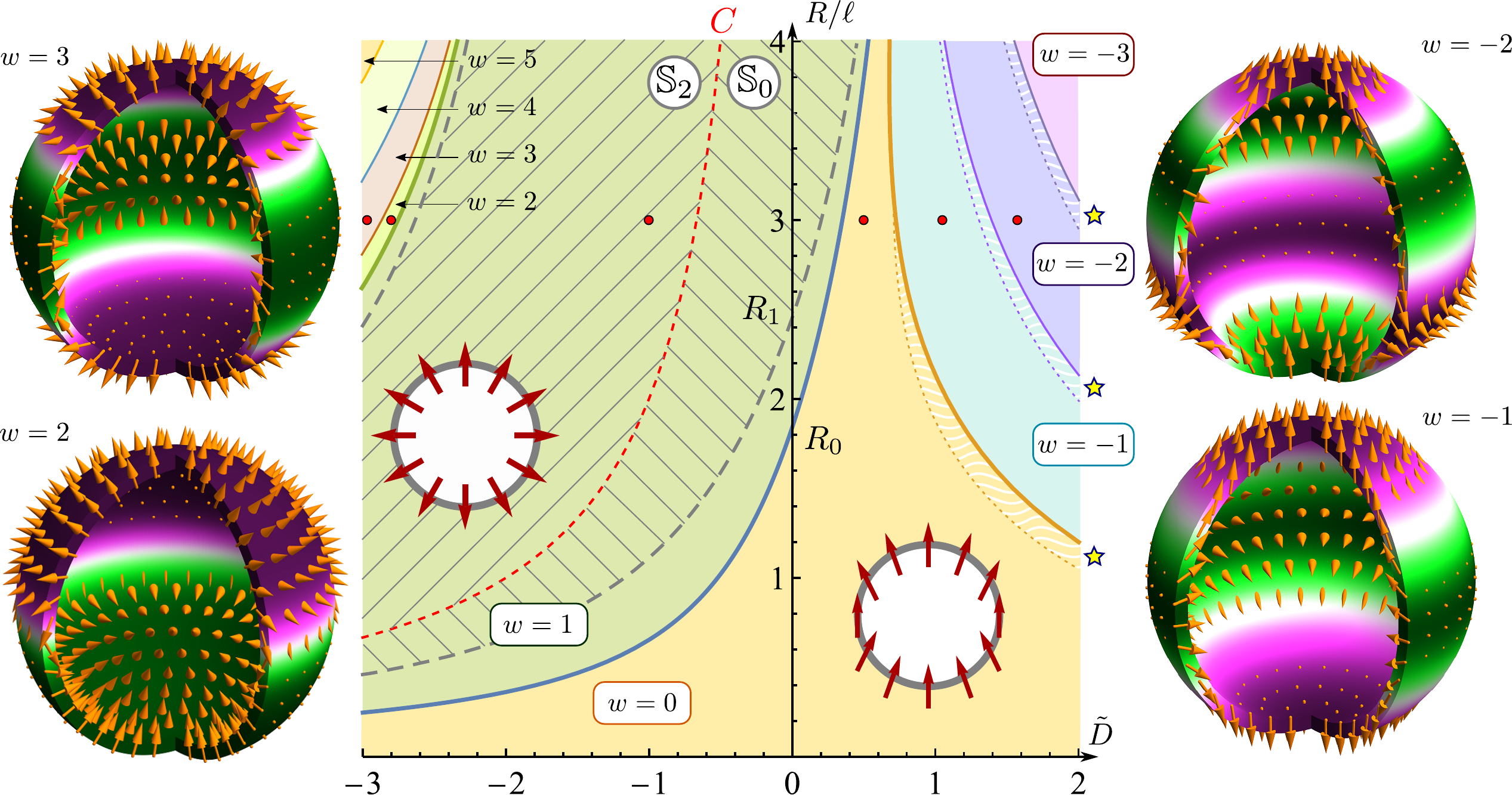}
	\caption{Diagram of magnetic ground states of a spherical shell\newtext{: an answer to the introductory question (ii).} The diagram was created by comparing the energies of different solutions of Eq.~\eqref{eq:Theta}. Even and odd helicity numbers $w$ correspond to $Q=0$ and $|Q|=1$ respectively. Each state with a given $w$ is doubly degenerate with respect to the transformation $\vec{m}\to-\vec{m}$, which results in $Q\to-Q$.  Small red dots correspond to the magnetization distributions \newtext{obtained by micromagnetic simulations and} shown as schematic insets ($w=0,1$) or rendered visualizations ($w=-1,-2,2,3$). Although skyrmion solutions do not form the ground state of the system, skyrmions with $w=0$ and $w=2$ can exist as topologically stable excitations in the dashed areas $\mathbb{S}_0$ and $\mathbb{S}_2$, respectively. Line $C$ is the line of skyrmion collapse, it is determined by the condition $D=-D_c$. Critical sphere radii $R_0$ and $R_1$ are explained in Fig.~\ref{fig:bifurcation}. \newtext{Star markers denote regions, where the elliptical instability of the given state can occur.} The other notations are the same as in Fig.~\ref{fig:Q0}}\label{fig:Gr-States}
\end{figure*}

\section{Case of a spherical shell}
As the simplest example we consider a thin spherical shell with radius $R$. For the case of easy-normal anisotropy ($K>0$) there exists a class of azimuthally symmetric solutions  $\vec{m}=\vec{e}_\vartheta\sin\theta+\vec{n}\cos\theta$, see Appendix~\ref{app:spherical_shell}.  The basis vector $\vec{e}_\vartheta$ points, tangential to the surface, towards the direction of increasing polar angle $\vartheta$ and $\vec{n}$ is the outward normal. The function $\theta=\theta(\vartheta)$ satisfies the following equation
\begin{equation}\label{eq:Theta}
\begin{split}
\theta''+\cot\vartheta\theta'-&\sin\theta\cos\theta\left[\frac{\cos2\vartheta}{\sin^2\vartheta}+\frac{R^2}{\ell^2}-\frac{4D}{D_c}\right]\\
+&2\cot\vartheta\sin^2\theta\left(1+\frac{D}{D_c}\right)=0.
\end{split}
\end{equation}
Here, $\ell=\sqrt{A/K}$ is the characteristic magnetic length and $D_c=2A/R$ is the strength of the curvature-induced effective DMI that is solely exchange-driven.\cite{Sheka15} This geometrical DMI contribution competes with the intrinsic spin-orbit-driven DMI.  Full compensation takes place when $D=-D_c$.

\newtext{In the limit $R\to\infty$ the equation \eqref{eq:Theta} is transformed\footnote{Before applying the limit $R\to\infty$ one should make a change of independent variable $\vartheta=\rho/R$, where $\rho$ is a distance along a meridian direction.} into the standard equation for chiral skyrmions in a planar film.\cite{Bogdanov99,Leonov16} This enables us to use the term ``skyrmion'' for a localized solution of the equation \eqref{eq:Theta}.}

There are two kinds of boundary conditions (BC) possible for Eq.~\eqref{eq:Theta}, namely (i) $\theta(0)=0$, $\theta(\pi)=(w-1)\pi$, and (ii)  $\theta(0)=\pi$, $\theta(\pi)=w\pi$. Here, $w\in\mathbb{Z}$, the helicity number, is formally a winding number of the magnetization along a circle loop passing through both pole points $\vartheta=0$ and $\vartheta=\pi$. Using the helicity number one can introduce the chirality of the structure: $\mathcal{C}=\mathrm{sgn}(w-1)$. Thus, the skyrmions shown in Fig.~\ref{fig:Q0}A and A$'$ have the chiralities $\mathcal{C}=-1$ and $\mathcal{C}=+1$, respectively.

From the general expression for the gyrocoupling vector \eqref{eq:gyro-local}, it follows that the mapping index for an azimuthally symmetrical solution $\vec{m}=\vec{m}(\vartheta)$ is
\begin{equation}\label{eq:Q-symm}
Q=-\frac{1}{2}\left[(\vec{m}\cdot\vec{n})|_{\vartheta=0}+(\vec{m}\cdot\vec{n})|_{\vartheta=\pi}\right]
\end{equation}
which implies that $Q=0,\pm1$ for the mentioned class of solutions. It is interesting to note that a one-dimensional magnetization in the planar case, $\vec{m}=\vec{m}(x)$, results in $Q=0$. However, in the case of a spherical shell a solution with $|Q|=1$ is possible even if $\vec{m}$ depends on one coordinate only.
According to \eqref{eq:Q-symm} an even $w$ results in $Q=0$ for both kinds of BC and an odd $w$ results in $Q=-1$ and $Q=+1$ for BC of type (i) and (ii), respectively. Note that, in contrast to the mapping degree $Q$, \emph{the helicity number $w$ is not a topological invariant}:  we merely use it for the classification of solutions. Any two solutions with different $w$ but with the same $Q$ belong to the same homotopy class and they can be transformed into each other by means of a continuous deformation of the vector field $\vec{m}$.

Any two solutions of \eqref{eq:Theta}, obtained under different kinds of BC but for the same $w$, differ by sign only: $\vec{m}\to-\vec{m}$. The latter transformation does not change the energy of the system, as the energy functional $E$, in the absence of external fields, is quadratic with respect to components of vector $\vec{m}$. However, it changes the sign of $Q$ because the mapping Jacobian $J$ is cubic in the magnetization. Thus a state with given $w$ is doubly degenerate with respect to replacements $\vec{m}\to-\vec{m}$ and $Q\to-Q$.


\subsection{Topologically trivial case $Q=0$: skyrmion solutions}\label{sbsec:Q0}

Equation \eqref{eq:Theta} can have skyrmion solutions for the cases $w=0$ and $w=2$, see Fig.~\ref{fig:Q0}A and Fig.~\ref{fig:Q0}A$'$, respectively.
In contrast to the planar case, where the skyrmion solution has $|Q|=1$, on a spherical shell the skyrmion is topologically trivial ($Q=0$).  Let us define the skyrmion radius as $R_s=R\sin\vartheta_s$, where $m_n(\vartheta_s)=0$. For the planar case $m_n=m_n(\rho)$ with $\rho$ being distance to the skyrmion center and $R_s$ can be defined analogously: $m_n(R_s)=0$. In planar films skyrmions are widely studied; it is well known\cite{Bogdanov94,Kiselev11,Rohart13} that the skyrmion radius strongly depends on the DMI constant $D$: the skyrmion collapses, $R_s\to0$, when $D\to0$; and $R_s\to\infty$ when $D\to D_0=\frac{4}{\pi}\sqrt{AK}$, see dashed line in Fig.~\ref{fig:Q0}. For this type of DMI, the so called hedgehog (N\'{e}el) skyrmions appear with zero azimuthal magnetization component, see Fig.~\ref{fig:Q0}B,B$'$. In planar films such type of skyrmions have been predicted theoretically\cite{Rohart13,Sampaio13} and were observed experimentally.\cite{Romming13,Romming15,Buettner15} The same type of skyrmions appear on a spherical shell with an analogous dependence $R_s=R_s(D)$, see the solid line in Fig.~\ref{fig:Q0}.  There are, however, a number of new, important features:

(i) Skyrmions collapse for a finite value of the DMI constant, $D=-D_c$, and as a consequence a skyrmion of finite radius exists for the case $D=0$, see point A in Fig.~\ref{fig:Q0} and the corresponding inset. The shift along the $D$ axis is due to the additional curvature-induced DMI \eqref{eq:Ed-eff}, which appears as an effective term in the exchange interaction.\cite{Sheka15}

(ii) For a given radius $R$ of the spherical shell the skyrmion exists for a certain range of the DMI constant, $D_l(R)<D<D_r(R)$, see Fig.~\ref{fig:Q0} and Fig.~\ref{fig:Gr-States}. Beyond this range, at $R_s=R$, see Fig.~\ref{fig:Q0}, the skyrmion transforms into the 3D-onion state.

(iii) In contrast to the planar case where the function $R_s(D)$ is even, the corresponding curve for spherical shells is highly asymmetrical.

\newtext{The analytically obtained dependence $R_s(D)$ agrees well with micromagnetic simulations data shown by open squares markers in Fig.~\ref{fig:Q0}, for details see Appendix~\ref{app:simuls}.}


Fig.~\ref{fig:Q0} shows possible equilibrium states for the case $Q=0$, answering our introductory question (i) about the physically stable magnetization structures in topologically different sectors.

\subsection{Diagram of ground states}\label{sbsec:ground-states}
 Though a continuous transition between solutions with $Q=0$ and $|Q|=1$ is not possible (topological stability), a solution with $|Q|=1$ can have lower energy than the corresponding solution with $Q=0$ for some range of parameters. In order to clarify this picture and answer question (ii) about the globally stable magnetization configurations, we build the diagram of the ground states for the class of azimuthally symmetrical solutions determined by Eq.~\eqref{eq:Theta}, see Fig.~\ref{fig:Gr-States}. One can distinguish two main states: the hedgehog states with $w=1$ ($|Q|=1$) and 3D-onion state with $w=0$ ($Q=0$). However, for large enough sphere radii and magnitudes of the DMI constant a variety of states with higher helicity numbers appear. These states can be interpreted as helical structures on a spherical shell. Similar skyrmionic structures were recently observed in disk-shaped chiral nanomagnets.\cite{Streubel15,Beg15}

Like the solitary skyrmion on a planar film, the skyrmion on a spherical shell does not form the magnetic ground state, yet skyrmions with $w=0$ and $w=2$ can exist as topologically stable excitations, see the dashed areas $\mathbb{S}_0$ and $\mathbb{S}_2$, respectively.

\newtext{The diagram  of the ground states (Fig.~\ref{fig:Gr-States}) was built for the class of azimuthally symmetrical solutions $\vec{m}=\vec{m}(\vartheta)$. Hence, we address the question about azimuthal stability of these solutions. Performing a standard stability analysis, see Appendix~\ref{app:stability}, we found a number of narrow regions, where elliptical instability\cite{Bogdanov94a} is possible. Remarkably, the instability regions are in the vicinity of boundaries which separate different magnetization states, see Fig.~\ref{fig:Gr-States}.}
\begin{figure}
	\begin{center}
	\includegraphics[scale=0.6]{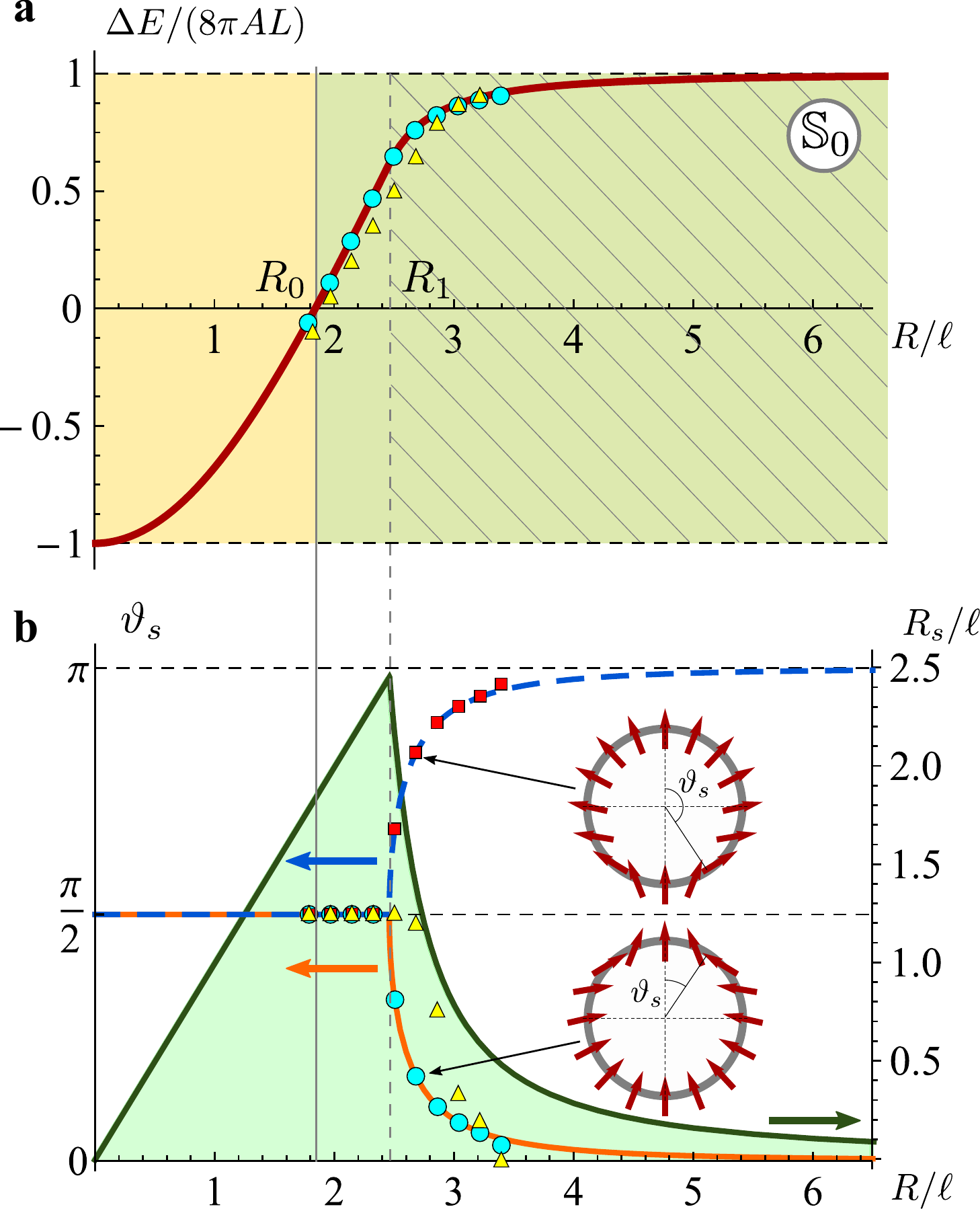}
	\end{center}
	\caption{Formation of the skyrmion state for the case $D=0$. Inset (a) shows the dependence of the energy difference $\Delta E=E_{w=0}-E_{w=1}$ of the solution $\theta(\vartheta)$ for the case $w=0$ and the hedgehog solution $\theta=0,\,\pi$, when $w=1$. The critical radius $R_0$ separates two phases of the ground state: 3D-onion state for $R<R_0$ and hedgehog state for $R>R_0$, see Fig.~\ref{fig:Gr-States}. For $R=R_1$ the 3D-onion state experiences instability, which results in the skyrmion formation. Skyrmion exists as a topologically stable excitation of the hedgehog state for the case $R>R_1$, see also region $\mathbb{S}_0$ in Fig.~\ref{fig:Gr-States}. Inset (b) demonstrates the dependence of the skyrmion radius (angular $\vartheta_s$ as well as lateral $R_s$) on the sphere radius $R$. Symbols correspond to the results of micromagnetic simulations: disks and squares -- the magnetostatic interaction is reduced to the easy-surface anisotropy, triangles --  the full scale simulations with magnetostatics is included.}\label{fig:bifurcation}
\end{figure}

\section{Skyrmion formation without DMI}

\begin{figure}
	\begin{center}
	\includegraphics[scale=0.6]{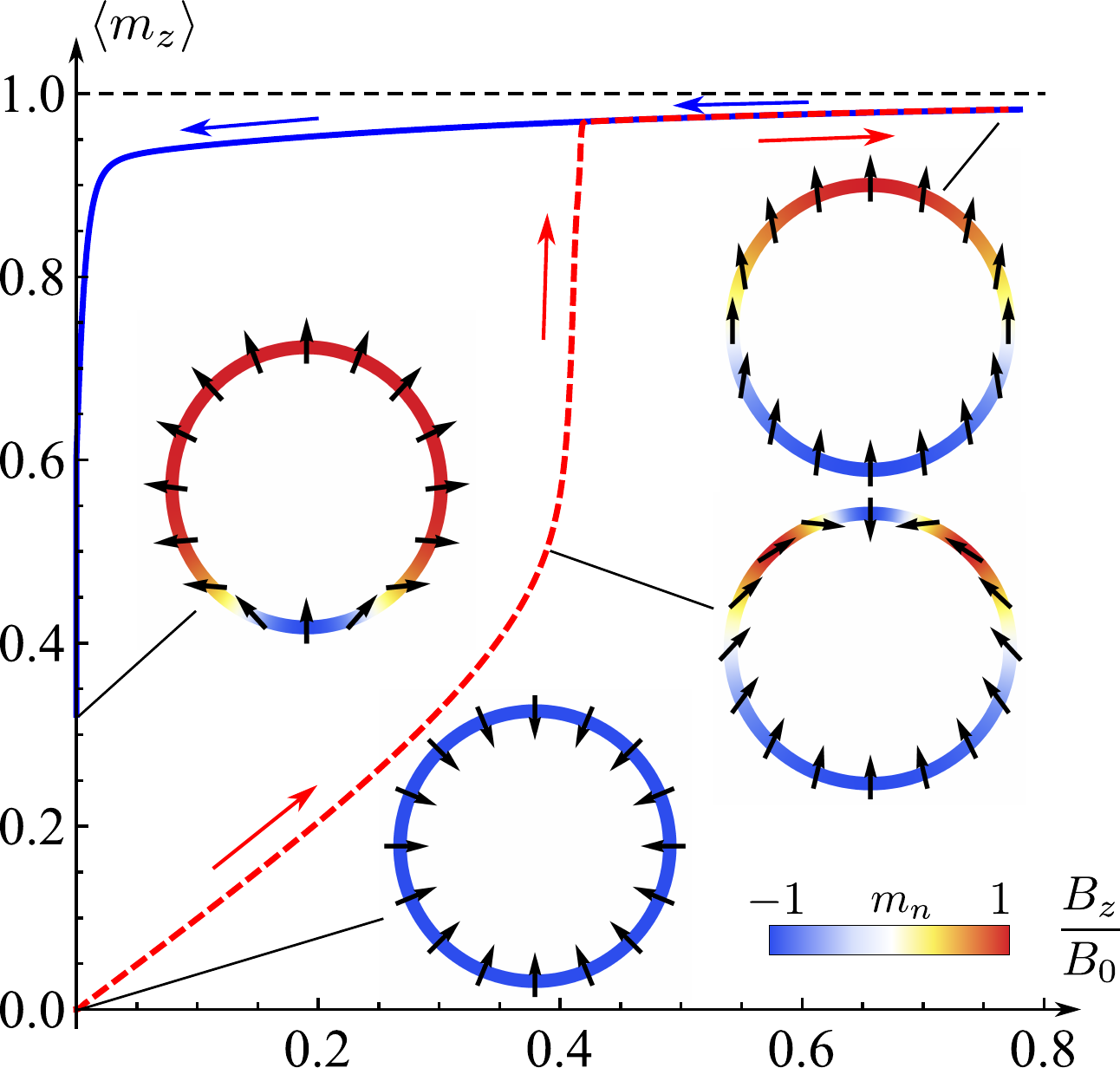}
	\end{center}
    \caption{Skyrmion formation by means of a uniform magnetic field (results of micromagnetic simulations). The directed lines show the magnetic hysteresis loop of the $z$ component of $m$ in an external magnetic field along the $z$ axis.  The arrows represent the increase (red/dashed) and subsequent decrease (blue/solid) of the magnetic field.  The field is normalized by the saturation field $B_0=4\pi M_s$. Insets, which show the vertical cross-section of the spherical shell, trace the formation of a Skyrmion.}\label{fig:hyst}
\end{figure}

The results on static skyrmion state configurations on spherical shells immediately pose the problem whether and how these states can be realized. In the following, we discuss in more detail an intriguing case of the skyrmion formation, when $D=0$. For this purpose, we will move along the vertical axis of the ground states diagram (Fig.~\ref{fig:Gr-States}) starting from small sphere radii $R$. When the sphere's radius is small enough ($R<R_0\approx1.842\ell$), the ground state of the system is topologically trivial ($Q=0$) and close to the uniform state, we call it a 3D-onion state, see Fig.~\ref{fig:Gr-States} and Fig.~\ref{fig:bifurcation}a. The ground state of the spheres with $R>R_0$ is one of the hedgehog states with $Q=\pm1$. Due to the topological stability the 3D-onion state survives when the sphere radii $R_0<R<R_1\approx2.458\ell$. At the point $R=R_1$ the 3D-onion state becomes unstable resulting in a fork-like bifurcation, see Fig.~\ref{fig:bifurcation}b, where the lines are obtained by solving Eq.~\ref{eq:Theta} with the boundary conditions $\theta(0)=0$, $\theta(\pi)=-\pi$ and dots correspond to the micromagnetic simulations. As a result of the bifurcation a skyrmion is formed either on the north or on the south pole of the sphere, see Fig.~\ref{fig:bifurcation}. The skyrmion exists as a topologically stable excitation of a hedgehog state for the case $R>R_1$, see region $\mathbb{S}_0$ in Figs.~\ref{fig:Gr-States},~\ref{fig:bifurcation}a. However the radius of the skyrmion $R_s$ decreases rapidly when the radius of the sphere further increases, see Fig.~\ref{fig:bifurcation}b.

The observed behavior of this system can be explained as follows. The 3D-onion state can be interpreted as a skyrmion solution with the angular radius $\vartheta_s=\pi/2$, or with lateral radius $R_s=R$. However, it is well known\cite{Bogdanov94,Kiselev11,Rohart13} that skyrmions on a planar film collapse when the intrinsic spin-orbit-driven DMI vanishes, i.e., in the limit $D=0$. Thus, it is natural to expect the skyrmion to collapse for the case $R\to\infty$ when the curvature effect vanishes. That is why the instability of the 3D-onion state appears for a certain value of $R=R_1$ and with a further increase of the radius of the sphere the skyrmion collapses to either the north or the south pole, see Fig.~\ref{fig:bifurcation}b. However, for a sphere radius $R\gtrapprox R_1$ skyrmions with finite radii exist, see Fig.~\ref{fig:bifurcation}b. This can be interpreted as a skyrmion stabilization due to the curvature-induced, exchange-driven effective DMI.\cite{Gaididei14,Sheka15} The obtained results agree very well with micromagnetic simulations based on a model where the magnetostatic interaction was reduced to the easy-surface anisotropy, see disk- and square-shaped symbols in Fig.~\ref{fig:bifurcation}b. It is important that taking into account the magnetostatic interaction does not change the physical picture, but results in an increase of the skyrmion radius, see triangle-shaped markers in Fig.~\ref{fig:bifurcation}b. The increase of the skyrmion radius appears due to the volume magnetostatic charges. Thus, the full micromagnetic simulation results vindicate our analytical approach that  neglect the dipolar stray-fields to construct the diagrams of equilibrium states.

It is noteworthy that the energy difference $\Delta E=E_{w=0}-E_{w=1}$ of the skyrmion (3D-onion) and the hedgehog solutions varies in the region $-E_{hg}^0<\Delta E<E_{BP}$, where $E_{hg}^0$ is the energy of the hedgehog state sphere with $R\to0$, and $E_{BP}$ is the energy of the Belavin-Polyakov soliton,\cite{Belavin75} which is equal to the energy of a skyrmion with infinitesimal radius, see Fig.~\ref{fig:bifurcation}a. It is also remarkable that $E_{hg}^0=E_{BP}=8\pi A L$, which matches the energy of the vortex-antivortex pair with the opposite polarities just before their annihilation.\cite{Tretiakov07,Hertel06}

Finally, we demonstrate how the skyrmion configuration can be created by means of a uniform magnetic field. We consider a spherical shell, whose radius $R\gtrapprox R_1$ corresponds to the hedgehog ground state. By means of micromagnetic simulations (see Appendix~\ref{app:simuls}) we find that the adiabatically slow increase of the external uniform magnetic field results in the transition from the hedgehog state to the 3D-onion state, see Fig.~\ref{fig:hyst}. A subsequent decrease of the field leads to a skyrmion state formation.  This transition from $Q=1$ to $Q=0$ is topologically forbidden for a continuous system and appears here merely due to the discretization.  However, since real magnetic crystals have a discrete structure, one can expect this behavior in strong enough external magnetic fields. This mechanism of the skyrmion formation is similar to the formation of onion magnetic states in nanorings.\cite{Rothman01}

\section{Conclusions}

In conclusion, we demonstrate that different types of axially symmetrical solutions of the magnetization  $\vec{m}=\vec{m}(\vartheta)$ exist for a thin ferromagnetic spherical shell. These solutions can be divided into three homotopic classes with topological index $Q=0,\,\pm1$.  To calculate $Q$ we developed the general expression for the mapping Jacobian \eqref{eq:gyro-local} valid for an arbitrary curvilinear shell. Skyrmion solutions are found in the topologically trivial class with $Q=0$. Remarkably, a skyrmion solution on a spherical shell can be stabilized by curvature effects only, namely by the curvature-induced, exchange-driven effective DMI.\cite{Gaididei14,Sheka15} This is in contrast to the planar case, where the spin-orbit-driven intrinsic DMI is required for the skyrmion stabilization.\cite{Bogdanov94,Kiselev11,Rohart13}  Since a skyrmion on a spherical shell is homotopic to a uniformly magnetized sphere, it can be induced by a strong uniform external magnetic field.

\newtext{Experimental advances in fabrications of curvilinear nanomagnets\cite{Streubel16a} make us optimistic in forthcoming experimental confirmation of the curvature stabilized skyrmions. Indeed, magnetic spherical nanoshells can be prepared\cite{Zhang09a,Cabot09,Gong14} by coating of a nonmagnetic spherical core with a ferromagnetic material. The small size (10--20~nm) of the obtained particles\cite{Zhang09a,Cabot09,Gong14} enable one to expect  discernible curvature effects. Note that spherical magnetic nanocaps with normally oriented anisotropy axis can also be created experimentally.\cite{Albrecht05,Ulbrich06,Makarov08,Makarov09} }

\section*{Acknowledgements}

V.P.K.\ acknowledges the Alexander von Humboldt Foundation for the support and IFW Dresden for kind hospitality. This work was funded in part by the European Research Council under the European Union's Seventh Framework Programme (FP7/2007-2013)/ERC grant agreement No.\ 306277 and the European Union Future and Emerging Technologies Programme (FET-Open Grant No.\ 618083). D.D.S.\ acknowledges Prof.\ Avadh Saxena for fruitful discussions.

\appendix
\section{Introduction of the curvilinear basis}\label{app:basis}
In order to formalize the geometry of the shell we use the parametric representation $\vec{\mathcal{R}}(\xi_1,\xi_2,\eta)=\vec{\varrho}(\xi_1,\xi_2)+\eta\vec{n}(\xi_1,\xi_2)$.
Here, $\vec{\varrho}  = \varrho_i \hat{\vec{x}}_i$ is the 3D position vector, which determines a 2D surface $\mathcal{S}$ embedded in $\mathbb{R}^3$ with $\xi_\alpha$ being local curvilinear coordinates on $\mathcal{S}$.  The unit vector $\vec{n}$ denotes the surface normal and the parameter $\eta\in[-L/2,L/2]$ is the corresponding curvilinear coordinate along the normal direction. We restrict ourselves to the limiting case $L\to0$. Specifically, we assume that the thickness $L$ is much smaller than the curvature radius as well as the characteristic magnetic length $\ell$. As a consequence, we assume that the magnetization is uniform along the normal direction: $\vec{m}=\vec{m}(\xi_1,\xi_2)$.

The parameterization $\vec{\varrho}=\vec{\varrho}(\xi_1,\xi_2)$ induces the natural tangential basis $\vec g_\alpha=\partial_\alpha\vec{\varrho}$ with the corresponding metric tensor $g_{\alpha\beta}=\vec{g}_\alpha\cdot\vec{g}_{\beta}$. Assuming that vectors $\vec g_{\alpha}$ are orthogonal, one can introduce the orthonormal basis $\{\vec{e}_1,\,\vec{e}_2,\,\vec n\}$, where $\vec{e}_\alpha=\vec{g}_\alpha/\sqrt{g_{\alpha\alpha}}$ and $\vec{n}=\vec{e}_1\times\vec{e}_2$.
Using the Gau{\ss}-Codazzi formula and Weingarten's equations\cite{Dubrovin84p1} one can obtain the following differential properties of the basis vectors
\begin{equation}\label{eq:basis-diff}
\begin{split}
&\nabla_{\!\!\alpha}\vec{e}_\beta=h_{\alpha\beta}\vec{n}-\Omega_\alpha\epsilon_{\beta\gamma}\vec{e}_\gamma,\\
&\nabla_{\!\!\alpha}\vec{n}=-h_{\alpha\beta}\vec{e}_\beta.
\end{split}
\end{equation}
Here, matrix $||h_{\alpha\beta}||=-||\vec{e}_\beta\cdot(\vec{e}_\alpha\cdot\vec{\nabla})\vec{n}||$ is a tensor, known as the Weingarten map or modified second fundamental form.\cite{Kamien02} The formula $||h_{\alpha\beta}||=||b_{\alpha\beta}/\sqrt{g_{\alpha\alpha}g_{\beta\beta}}||$ with $b_{\alpha\beta} = {\vec{n}}\cdot \partial_\beta\vec{g}_\alpha$ being components of the second fundamental form, is practically useful. The Weingarten map determines the Gau{\ss} curvature $\mathcal{K}=\det||h_{\alpha\beta}||$ and the mean curvature $\mathcal{H}=\mathrm{tr}||h_{\alpha\beta}||$.  Components of the spin connection vector $\vec\Omega$ are determined by the relation $\Omega_\gamma = \frac12 \epsilon_{\alpha\beta}\vec{e}_\alpha\cdot \nabla_{\!\!\gamma} \vec{e}_\beta$.

In general, the basis $\vec g_{\alpha}$ need not be orthogonal. However, if the vectors $\vec g_1$ and $\vec g_2$ are not collinear then one can always introduce the orthonormal basis $\{\vec{e}_1,\,\tilde{\vec{e}}_2,\,\vec n\}$ in the following way: $\vec{e}_1=\vec{g}_1/\sqrt{g_{11}}$, $\tilde{\vec{e}}_2=\vec{g}^2/\sqrt{g^{22}}$, and $\vec{n}=\vec{e}_1\times\tilde{\vec{e}}_2$, where $\vec{g}^\alpha=g^{\alpha\beta}\vec{g}_\beta$ is dual tangential basis with $||g^{\alpha\beta}||=||g_{\alpha\beta}||^{-1}$. However, in this case the relations \eqref{eq:basis-diff} should be revised.

\section{Gyrocoupling vector in a curvilinear reference frame}\label{app:gyrovector}
We start from the Cartesian representation $\mathrm{J}_i=-\epsilon_{ijk}\vec{m}\cdot[\partial_j\vec m\times\partial_k\vec m]/2$. Taking into account that the magnetization is uniform along the shell thickness, i.e.\ the normal coordinate, one can replace the del operator by its surface analogue $\vec{\nabla}_{\!\textsc{s}}=\vec{e}_\alpha\nabla_{\!\!\alpha}$ and represent the Cartesian derivatives via its curvilinear counterparts $\partial_i=(\hat{\vec x}_i\cdot\vec{e}_\alpha)\nabla_{\!\!\alpha}$.
Simple calculations result in $\vec{\mathrm{J}}=\mathcal{J}\vec{n}$, where $\mathcal{J}$ is the mapping Jacobian defined in the main text. Introducing the curvilinear magnetization components $\vec{m}=m_\alpha\vec{e}_\alpha+m_n\vec{n}$ and taking into account differential properties of the curvilinear basis \eqref{eq:basis-diff} one obtains
\begin{equation}\label{eq:qn-ma-mn}
\begin{split}
\mathcal{J}&=-\epsilon_{\alpha\beta}\epsilon_{\gamma\delta}\Bigl[m_\beta(\nabla_{\!\!\delta} m_\alpha)(\nabla_{\!\!\gamma} m_n)+\frac{m_n}{2}(\nabla_{\!\!\gamma} m_\alpha)(\nabla_{\!\!\delta} m_\beta)\\
&+h_{\alpha\gamma}m_n(m_\delta\nabla_{\!\!\beta} m_n-m_n\nabla_{\!\!\delta} m_\beta)+h_{\alpha\nu}m_\nu m_\delta\nabla_{\!\!\beta}  m_\gamma\Bigr]\\
&+\epsilon_{\alpha\beta}\Omega_\beta m_\gamma\left[m_n\nabla_{\!\!\alpha} m_\gamma-m_\gamma\nabla_{\!\!\alpha} m_n-h_{\alpha\gamma}\right]-\mathcal{K}m_n.
\end{split}
\end{equation}
When deriving \eqref{eq:qn-ma-mn} the constraint $|\vec{m}|=1$ is obeyed.  Substituting the angular representation for the magnetization components into \eqref{eq:qn-ma-mn}, one obtains formula \eqref{eq:gyro-local}.

Notice the difference between volume $\vec{\mathrm{J}}$ and surface  $\vec{\mathcal{J}}$ gyrocoupling vectors. $\vec{\mathrm{J}}$ can be introduced for a vector field $\vec{m}$ defined in a 3D region $\mathfrak{D}$. One can cut out from $\mathfrak{D}$ a curvilinear shell with small but finite thickness $L$. If $\vec{m}$ is \emph{uniform along the shell thickness} (otherwise $\vec{\mathcal{J}}$ can not be introduced), then $\vec{\mathrm{J}}=\vec{\mathcal{J}}$. However, if we build a 2D  surface $\mathcal{S}$ ($L=0$) in $\mathfrak{D}$, then $\vec{\mathrm{J}}\ne\vec{\mathcal{J}}$ on the surface.

\section{Magnetic interactions on a curvilinear shell}\label{app:interactions}
Let us first consider the exchange interaction. Under the main assumption of magnetization uniformity along the shell thickness, one can represent the exchange energy density in its curvilinear form,
\begin{subequations}\label{eq:Eex-detail}
	\begin{align}
\mathscr{E}_{\mathrm{ex}}=&\nabla_{\!\!\alpha}\vec{m}\cdot\nabla_{\!\!\alpha}\vec{m}=\mathscr{E}_{\mathrm{ex}}^0+\mathscr{E}_{\mathrm{ex}}^\textsc{d}+\mathscr{E}_{\mathrm{ex}}^\textsc{a},\\
\mathscr{E}_{\mathrm{ex}}^0=&\nabla_{\!\!\alpha} m_\beta\nabla_{\!\!\alpha} m_\beta+\nabla_{\!\!\alpha} m_n\nabla_{\!\!\alpha} m_n,\\
\label{eq:Ed-eff}\mathscr{E}_{\mathrm{ex}}^\textsc{d}=&2h_{\alpha\beta}\left(m_\beta\nabla_{\!\!\alpha} m_n-m_n\nabla_{\!\!\alpha} m_\beta\right)\\ \nonumber
&+2\epsilon_{\alpha\beta}\Omega_\gamma m_\beta\nabla_{\!\!\gamma} m_\alpha,\\
\mathscr{E}_{\mathrm{ex}}^\textsc{a}=&(h_{\alpha\gamma}h_{\gamma\beta}+\Omega^2\delta_{\alpha\beta})m_\alpha m_\beta+(\mathcal{H}^2-2\mathcal{K})m_n^2\\ \nonumber
&+2\epsilon_{\alpha\gamma}h_{\gamma\beta}\Omega_\beta m_\alpha m_n.
\end{align}
\end{subequations}
Here, $\mathscr{E}_{\mathrm{ex}}^0$ is the ``common'' isotropic exchange, $\mathscr{E}_{\mathrm{ex}}^\textsc{d}$ and $\mathscr{E}_{\mathrm{ex}}^\textsc{a}$ can be treated as an effective curvature-induced DMI and anisotropy, respectively.
In the angular representation the exchange energy reads\cite{Gaididei14}
\begin{equation}\label{eq:Eex} \mathscr{E}_{\mathrm{ex}}=\left[\vec{\nabla}\theta-\vec{\Gamma}\right]^2+\left[\sin\theta(\vec{\nabla}\phi-\vec\Omega)-\cos\theta\,\partial_\phi\vec{\Gamma}\right]^2.
\end{equation}
Applying the same procedure for the DMI energy density $\mathscr{E}_{\textsc{d}}=m_n\partial_im_i-m_i\partial_im_n$ one obtains
\begin{subequations}\label{eq:Ed-detail}
	\begin{align}
		\mathscr{E}_{\textsc{d}}=&\mathscr{E}_{\textsc{d}}^0+\mathscr{E}_{\textsc{d}}^\textsc{a},\\
		\mathscr{E}_{\textsc{d}}^0=&m_n\nabla_{\!\!\alpha} m_\alpha-m_\alpha\nabla_{\!\!\alpha} m_n,\\
		\mathscr{E}_{\textsc{d}}^\textsc{a}=&-\epsilon_{\alpha\beta}\Omega_\beta m_\alpha m_n-\mathcal{H}m_n^2,
	\end{align}
\end{subequations}
where an additional term $\mathscr{E}_{\textsc{d}}^\textsc{a}$ can be interpreted as an effective curvature-induced anisotropy. Substituting the angular parameterization into  \eqref{eq:Ed-detail} results in the expression \eqref{eq:Ed}.

In the angular representation the density of the anisotropy energy looks particularly simple $\mathscr{E}_{\mathrm{a}}=-K(\vec{m}\cdot\vec{n})^2=-K\cos^2\theta$, because the anisotropy has the symmetry of the surface.

\section{Case of a spherical shell}\label{app:spherical_shell}
 In order to describe a spherical shell of radius $R$ we use the parameterization
$\vec{\varrho}(\vartheta,\chi)=R\left(\sin\vartheta\cos\chi\,\hat{\vec{x}}+\sin\vartheta\sin\chi\,\hat{\vec{y}}+\cos\vartheta\,\hat{\vec{z}}\right)$.
Here, $\vartheta=\xi_1\in[0,\pi]$ and $\chi=\xi_2\in S^1$ are polar and azimuthal spherical angles, respectively. Basis vectors are $\vec{e}_1=\vec{e}_\vartheta$, $\vec{e}_2=\vec{e}_\chi$ and the normal vector $\vec{n}=\vec{e}_\vartheta\times\vec{e}_\chi$ is directed outward of the sphere.
In this case the Weingarten map is the diagonal matrix with components $h_{\alpha\beta}=-\delta_{\alpha\beta}/R$, and consequently $\vec\Gamma=-\vec{\varepsilon}/R$, $\mathcal{H}=-2/R$ and $\mathcal{K}=1/R^2$. The spin-connection vector has only one (azimuthal) component:  $\vec{\Omega}=-\vec{e}_\chi\cot\vartheta/R$.

Taking into account \eqref{eq:Eex} and \eqref{eq:Ed} one can show that in the case of easy-normal anisotropy ($K>0$) the  functions $\theta=\theta(\vartheta,\chi)$ and $\phi=\phi(\vartheta,\chi)$, which minimize the energy functional $E$, are a solution of the following Euler-Lagrange equations
\begin{subequations}\label{eq:theta-phi-sphere}
	\begin{align}
		\label{eq:theta-sphere}&\Delta\theta-\sin\theta\cos\theta\left[(\vec\nabla\phi-\vec{\Omega})^2+\frac{1}{\ell^2}-\frac{1}{R^2}\left(1+\frac{4D}{D_c}\right)\right]\\ \nonumber
		&+\frac{2}{R}\sin^2\theta\left(1+\frac{D}{D_c}\right)(\vec\nabla\phi-\vec{\Omega})\cdot\partial_\phi\vec{\varepsilon}=0, \\
		\label{eq:phi-sphere}&\vec{\nabla}\cdot\left[\sin^2\theta(\vec\nabla\phi-\vec{\Omega})\right]-\frac{2}{R}\sin^2\theta\left(1+\frac{D}{D_c}\right)\vec{\nabla}\theta\cdot\partial_\phi\vec{\varepsilon}=0.
	\end{align}
\end{subequations}
When deriving \eqref{eq:theta-phi-sphere} we used that for a spherical shell $\vec\Gamma=-\vec{\varepsilon}/R$ and $\mathcal{H}=-2/R$. In addition, we use the following general properties of vector $\vec{\varepsilon}$
\begin{equation}
\vec{\nabla}\cdot\vec{\varepsilon}=(\vec{\nabla}\phi-\vec\Omega)\cdot\partial_\phi\vec{\varepsilon},\quad \vec{\nabla}\cdot\partial_\phi\vec{\varepsilon}=-(\vec{\nabla}\phi-\vec\Omega)\cdot\vec{\varepsilon}
\end{equation}

Taking into account that the vector of spin connection is $\vec{\Omega}=\vec{e}_2\Omega_2(\vartheta)$ one can see that Eqs.~\eqref{eq:theta-phi-sphere} have a solution $\theta=\theta(\vartheta)$, $\phi=0,\,\pi$. In this case Eq.~\eqref{eq:phi-sphere} turns to identity and the function $\theta(\vartheta)$ can be determined as a solution of Eq.~\eqref{eq:theta-sphere}. Since $m_2\equiv0$  for the considered class of solutions, one can consider the colatitude angle $\theta=\theta(\vartheta)$ as the only parameter: $\vec{m}=\vec{e}_1\sin\theta(\vartheta)+\vec{n}\cos\theta(\vartheta)$. In this case the function $\theta(\vartheta)$ is determined by equation \eqref{eq:Theta}. For the axially symmetric solutions $\theta=\theta(\vartheta)$ and  $\phi=\phi(\vartheta)$ the gyrocoupling vector \eqref{eq:gyro-local} for the spherical shell can be written as follows
\begin{equation}
\vec{\mathcal{J}}=\frac{\vec{n}}{R^2\sin\vartheta}\frac{\mathrm{d}}{\mathrm{d}\vartheta}\left(\cos\theta\cos\vartheta-\sin\theta\sin\phi\sin\vartheta\right).
\end{equation}
The integration over the sphere $S^2$ results in the topological index \eqref{eq:Q-symm}.

\section{Stability analysis}\label{app:stability}
\newtext{Our goal is to analyze the stability of azimuthally symmetrical solutions, shown in Fig.~\ref{fig:Gr-States}. For this purpose we use the parameterization $\vec{m}=\vec{e}_\vartheta\sin\phi\sin\theta+\vec{e}_\chi\cos\phi+\vec{n}\sin\phi\cos\theta$. Taking into account the exchange \eqref{eq:Eex-detail}, DMI \eqref{eq:Ed-detail} and anisotropy, $\mathscr{E}_{\mathrm{a}}=-K\sin^2\phi\cos^2\theta$, contributions we can construct the energy functional $E=E[\theta,\,\phi]$. This functional has an extremal for $\phi_0=\pi/2$, $\theta_0=\theta_0(\vartheta)$, where the function $\theta_0(\vartheta)$ is a solution of Eq.~\eqref{eq:Theta}. We need to check whether the solution $\phi_0$ and $\theta_0$ corresponds to the energy minimum. For this purpose, we consider small deviations $\phi=\pi/2+\tilde{\phi}(\vartheta,\chi)$ and $\theta=\theta_0(\vartheta)+\tilde{\theta}(\vartheta,\chi)$.
Now the harmonic approximation of the energy reads
\begin{equation}\label{eq:E-harm}
E\approx E_0+AL\int\limits_{0}^{2\pi}\mathrm{d}\chi\int\limits_{0}^{\pi}\mathrm{d}\vartheta\sin\vartheta\,\vec{\psi}^\textsc{t}\hat{\mathcal{H}}\vec{\psi}.
\end{equation}
Here, $E_0=E[\theta_0,\phi_0]$ is stationary value of the energy functional, $\vec{\psi}=(\tilde{\phi},\tilde{\theta})^\textsc{t}$ and the operator $\hat{\mathcal{H}}$ reads
\begin{equation}
\label{eq:H}
\hat{\mathcal{H}}=\begin{pmatrix}
-\Delta+\mathcal{U}_1(\vartheta)&-\mathcal{W}(\vartheta)\partial_\chi\\
\mathcal{W}(\vartheta)\partial_\chi & -\Delta+\mathcal{U}_2(\vartheta)
\end{pmatrix},
\end{equation}
where $\Delta$ is the angular part of the Laplacian in the spherical reference frame. The potentials are as follows
\begin{equation}\label{eq:U12-W}
\begin{split}
\mathcal{U}_1=&-(\theta_0'+1)^2+\frac{\cos^2(\vartheta+\theta_0)}{\sin^2\vartheta}+\frac{R^2}{\ell^2}\cos^2\theta_0-\\
&-\frac{D}{D_c}\left[2(\theta_0'+1)+\Xi\right],\\
\mathcal{U}_2=&\frac{\cos2(\vartheta+\theta_0)}{\sin^2\vartheta}+\frac{R^2}{\ell^2}\cos2\theta_0-2\frac{D}{D_c}\Xi,\\
\mathcal{W}=&-2\frac{\cos(\vartheta+\theta_0)}{\sin^2\vartheta}+2\frac{D}{D_c}\frac{\sin\theta_0}{\sin\vartheta},
\end{split}
\end{equation}
where $\Xi=2\cos2\theta_0+\cot\vartheta\sin2\theta_0$. One can easily check that the Euler equations with respect to small deviations $\tilde{\phi}$ and $\tilde{\theta}$ (Jacobi equation) have the solutions $\tilde{\phi}=\sum_{\mu}f_\mu(\vartheta)\sin\mu\chi$ and $\tilde{\theta}=\sum_{\mu}g_\mu(\vartheta)\cos\mu\chi$, with $\mu\in\mathbb{Z}$. Introducing $\vec{\psi}_\mu=\left(f_\mu,g_\mu\right)^\textsc{t}$ one can present the energy \eqref{eq:E-harm} in the form
\begin{equation}\label{eq:E-harm-mu}
E\approx E_0+\pi AL\sum\limits_\mu\int\limits_{0}^{\pi}\mathrm{d}\vartheta\sin\vartheta\,\vec{\psi}_\mu^\textsc{t}\hat{\mathcal{H}}_{\mu}\vec{\psi}_\mu,
\end{equation}
where
\begin{equation}\label{eq:H-mu}
\hat{\mathcal{H}}_{\mu}=\begin{pmatrix}
-\Delta_\vartheta+\frac{\mu^2}{\sin^2\vartheta}+\mathcal{U}_1(\vartheta)&\mu\mathcal{W}(\vartheta)\\
\mu\mathcal{W}(\vartheta) & -\Delta_\vartheta+\frac{\mu^2}{\sin^2\vartheta}+\mathcal{U}_2(\vartheta)
\end{pmatrix}
\end{equation}
is a Hermitian operator in the space of functions $\vec{\psi}_\mu$ with the scalar product $\langle\vec{\psi}_\mu^{(1)},\vec{\psi}_\mu^{(2)}\rangle=\int_0^\pi\vec{\psi}_\mu^{(1)}\cdot\vec{\psi}_\mu^{(2)}\sin\vartheta\mathrm{d}\vartheta$. Here, $\Delta_\vartheta f=(\sin\vartheta)^{-1}\partial_\vartheta(\sin\vartheta\partial_\vartheta f)$.}

\newtext{The solution $\phi_0$ and $\theta_0$ minimizes the energy functional $E$ iff all eigenvalues of the operator $\hat{\mathcal{H}}_\mu$ are positive for all $\mu$. Note that sign of $\mu$ does not effect the eigenvalues of the operator \eqref{eq:H-mu}. For a given pair of parameters $(D,R)$ we found numerically a set of eigenvalues of operator \eqref{eq:H-mu} for the range $\mu=\overline{0,10}$ and for the fixed boundary conditions $\vec{\psi}_\mu(0)=\vec{\psi}_\mu(\pi)=\vec0$. Using the mentioned criterion, we found some narrow instability regions in the diagram Fig.~\ref{fig:Gr-States} in the vicinity of boundaries between states with different helicity numbers $w$.
These instability are found there only for modes $\mu=2$, thus following Ref.~\onlinecite{Bogdanov94a}, we call it elliptical instability.}

\section{Micromagnetic simulations}\label{app:simuls}
 In order to verify our analytical results we performed micromagnetic simulations with the \texttt{FinMag} code, which is the successor to the \texttt{Nmag} tool.\cite{Fischbacher07} We used the material parameters of cobalt: $A=1.6\times10^{-11}$ J/m, $M_s=1.1\times10^6$ A/m and $K=1.3\times10^6$ J/m$^3$, which are typical for Pt/Co/AlO$_x$ layer structures.\cite{Rohart13} For comparison, we also performed a simulation neglecting the magnetostatic interaction, assuming that it can be reduced to an effective easy-surface anisotropy, we used $K_{\mathrm{eff}}=5.1\times10^5$ J/m$^3$. These parameters correspond to $\ell=5.6$ nm. In all simulations the ratio $L/R=0.1$ is kept constant. The size of the discretization mesh is $0.1-0.2L$.

 \newtext{To verify the dependence $R_s(D)$ shown in Fig.~\ref{fig:Q0} we simulate the shell with radius $R=3\ell=16.8$~nm, thickness $L=0.3\ell=1.68$~nm and an average mesh size of 0.42~nm. Since the mesh discreteness breaks the topological stability, we are not able to obtain the skyrmions with small radii $R_s<0.2R$. The same geometrical parameters are used in simulations for Fig.~\ref{fig:Gr-States}.}

To simulate the formation of skyrmions by means of a uniform field, see Fig.~\ref{fig:hyst}, we consider a spherical shell with radius $R=15$~nm and thickness $L=1.5$~nm. The applied magnetic field is increased from zero up to the value 1.5 T with a rate of 330 mT/ns.  The field is then decreased back to zero with rate 82 mT/ns. In order to break the symmetry and avoid the unstable equilibrium state we introduce a radial $\vec{b}=b_0\vec{n}$ field with a small amplitude $b_0=5$ mT. Fields in  Fig.~\ref{fig:hyst} are normalized to the value $B_0=4\pi M_s=1.38$ T. In this numerical experiment we reduce magnetostatics to the effective easy-surface anisotropy.


\begin{thebibliography}{93}%
	\makeatletter
	\providecommand \@ifxundefined [1]{%
		\@ifx{#1\undefined}
	}%
	\providecommand \@ifnum [1]{%
		\ifnum #1\expandafter \@firstoftwo
		\else \expandafter \@secondoftwo
		\fi
	}%
	\providecommand \@ifx [1]{%
		\ifx #1\expandafter \@firstoftwo
		\else \expandafter \@secondoftwo
		\fi
	}%
	\providecommand \natexlab [1]{#1}%
	\providecommand \enquote  [1]{``#1''}%
	\providecommand \bibnamefont  [1]{#1}%
	\providecommand \bibfnamefont [1]{#1}%
	\providecommand \citenamefont [1]{#1}%
	\providecommand \href@noop [0]{\@secondoftwo}%
	\providecommand \href [0]{\begingroup \@sanitize@url \@href}%
	\providecommand \@href[1]{\@@startlink{#1}\@@href}%
	\providecommand \@@href[1]{\endgroup#1\@@endlink}%
	\providecommand \@sanitize@url [0]{\catcode `\\12\catcode `\$12\catcode
		`\&12\catcode `\#12\catcode `\^12\catcode `\_12\catcode `\%12\relax}%
	\providecommand \@@startlink[1]{}%
	\providecommand \@@endlink[0]{}%
	\providecommand \url  [0]{\begingroup\@sanitize@url \@url }%
	\providecommand \@url [1]{\endgroup\@href {#1}{\urlprefix }}%
	\providecommand \urlprefix  [0]{URL }%
	\providecommand \Eprint [0]{\href }%
	\providecommand \doibase [0]{http://dx.doi.org/}%
	\providecommand \selectlanguage [0]{\@gobble}%
	\providecommand \bibinfo  [0]{\@secondoftwo}%
	\providecommand \bibfield  [0]{\@secondoftwo}%
	\providecommand \translation [1]{[#1]}%
	\providecommand \BibitemOpen [0]{}%
	\providecommand \bibitemStop [0]{}%
	\providecommand \bibitemNoStop [0]{.\EOS\space}%
	\providecommand \EOS [0]{\spacefactor3000\relax}%
	\providecommand \BibitemShut  [1]{\csname bibitem#1\endcsname}%
	\let\auto@bib@innerbib\@empty
	\bibitem [{\citenamefont {Anderson}\ and\ \citenamefont
		{Toulouse}(1977)}]{Anderson77}%
	\BibitemOpen
	\bibfield  {author} {\bibinfo {author} {\bibfnamefont {P.~W.}\ \bibnamefont
			{Anderson}}\ and\ \bibinfo {author} {\bibfnamefont {G.}~\bibnamefont
			{Toulouse}},\ }\bibfield  {title} {\enquote {\bibinfo {title} {Phase slippage
				without vortex cores: Vortex textures in superfluid $^3$he},}\ }\href
	{\doibase 10.1103/physrevlett.38.508} {\bibfield  {journal} {\bibinfo
			{journal} {Phys. Rev. Lett.}\ }\textbf {\bibinfo {volume} {38}},\ \bibinfo
		{pages} {508--511} (\bibinfo {year} {1977})}\BibitemShut {NoStop}%
	\bibitem [{\citenamefont {Volovik}(2003)}]{Volovik03}%
	\BibitemOpen
	\bibfield  {author} {\bibinfo {author} {\bibfnamefont {G.}~\bibnamefont
			{Volovik}},\ }\href
	{http://www.zentralblatt-math.org/zmath/search/?an=01866310} {\emph {\bibinfo
			{title} {The universe in a {H}elium droplet}}}\ (\bibinfo  {publisher}
	{Oxford University Press},\ \bibinfo {address} {Oxford},\ \bibinfo {year}
	{2003})\BibitemShut {NoStop}%
	\bibitem [{\citenamefont {Hasan}\ and\ \citenamefont {Kane}(2010)}]{Hasan10}%
	\BibitemOpen
	\bibfield  {author} {\bibinfo {author} {\bibfnamefont {M.~Z.}\ \bibnamefont
			{Hasan}}\ and\ \bibinfo {author} {\bibfnamefont {C.~L.}\ \bibnamefont
			{Kane}},\ }\bibfield  {title} {\enquote {\bibinfo {title} {Colloquium :
				Topological insulators},}\ }\href {\doibase 10.1103/revmodphys.82.3045}
	{\bibfield  {journal} {\bibinfo  {journal} {Reviews of Modern Physics}\
		}\textbf {\bibinfo {volume} {82}},\ \bibinfo {pages} {3045--3067} (\bibinfo
		{year} {2010})}\BibitemShut {NoStop}%
	\bibitem [{\citenamefont {Moore}(2010)}]{Moore10a}%
	\BibitemOpen
	\bibfield  {author} {\bibinfo {author} {\bibfnamefont {Joel~E.}\ \bibnamefont
			{Moore}},\ }\bibfield  {title} {\enquote {\bibinfo {title} {The birth of
				topological insulators},}\ }\href {\doibase 10.1038/nature08916} {\bibfield
		{journal} {\bibinfo  {journal} {Nature}\ }\textbf {\bibinfo {volume} {464}},\
		\bibinfo {pages} {194--198} (\bibinfo {year} {2010})}\BibitemShut {NoStop}%
	\bibitem [{\citenamefont {Hsieh}\ \emph {et~al.}(2008)\citenamefont {Hsieh},
		\citenamefont {Qian}, \citenamefont {Wray}, \citenamefont {Xia},
		\citenamefont {Hor}, \citenamefont {Cava},\ and\ \citenamefont
		{Hasan}}]{Hsieh08}%
	\BibitemOpen
	\bibfield  {author} {\bibinfo {author} {\bibfnamefont {D.}~\bibnamefont
			{Hsieh}}, \bibinfo {author} {\bibfnamefont {D.}~\bibnamefont {Qian}},
		\bibinfo {author} {\bibfnamefont {L.}~\bibnamefont {Wray}}, \bibinfo {author}
		{\bibfnamefont {Y.}~\bibnamefont {Xia}}, \bibinfo {author} {\bibfnamefont
			{Y.~S.}\ \bibnamefont {Hor}}, \bibinfo {author} {\bibfnamefont {R.~J.}\
			\bibnamefont {Cava}}, \ and\ \bibinfo {author} {\bibfnamefont {M.~Z.}\
			\bibnamefont {Hasan}},\ }\bibfield  {title} {\enquote {\bibinfo {title} {A
				topological Dirac insulator in a quantum spin hall phase},}\ }\href {\doibase
		10.1038/nature06843} {\bibfield  {journal} {\bibinfo  {journal} {Nature}\
		}\textbf {\bibinfo {volume} {452}},\ \bibinfo {pages} {970--974} (\bibinfo
		{year} {2008})}\BibitemShut {NoStop}%
	\bibitem [{\citenamefont {Xu}\ \emph {et~al.}(2014)\citenamefont {Xu},
		\citenamefont {Alidoust}, \citenamefont {Belopolski}, \citenamefont
		{Richardella}, \citenamefont {Liu}, \citenamefont {Neupane}, \citenamefont
		{Bian}, \citenamefont {Huang}, \citenamefont {Sankar}, \citenamefont {Fang},
		\citenamefont {Dellabetta}, \citenamefont {Dai}, \citenamefont {Li},
		\citenamefont {Gilbert}, \citenamefont {Chou}, \citenamefont {Samarth},\ and\
		\citenamefont {Hasan}}]{Xu14}%
	\BibitemOpen
	\bibfield  {author} {\bibinfo {author} {\bibfnamefont {Su-Yang}\ \bibnamefont
			{Xu}}, \bibinfo {author} {\bibfnamefont {Nasser}\ \bibnamefont {Alidoust}},
		\bibinfo {author} {\bibfnamefont {Ilya}\ \bibnamefont {Belopolski}}, \bibinfo
		{author} {\bibfnamefont {Anthony}\ \bibnamefont {Richardella}}, \bibinfo
		{author} {\bibfnamefont {Chang}\ \bibnamefont {Liu}}, \bibinfo {author}
		{\bibfnamefont {Madhab}\ \bibnamefont {Neupane}}, \bibinfo {author}
		{\bibfnamefont {Guang}\ \bibnamefont {Bian}}, \bibinfo {author}
		{\bibfnamefont {Song-Hsun}\ \bibnamefont {Huang}}, \bibinfo {author}
		{\bibfnamefont {Raman}\ \bibnamefont {Sankar}}, \bibinfo {author}
		{\bibfnamefont {Chen}\ \bibnamefont {Fang}}, \bibinfo {author} {\bibfnamefont
			{Brian}\ \bibnamefont {Dellabetta}}, \bibinfo {author} {\bibfnamefont
			{Wenqing}\ \bibnamefont {Dai}}, \bibinfo {author} {\bibfnamefont
			{Qi}~\bibnamefont {Li}}, \bibinfo {author} {\bibfnamefont {Matthew~J.}\
			\bibnamefont {Gilbert}}, \bibinfo {author} {\bibfnamefont {Fangcheng}\
			\bibnamefont {Chou}}, \bibinfo {author} {\bibfnamefont {Nitin}\ \bibnamefont
			{Samarth}}, \ and\ \bibinfo {author} {\bibfnamefont {M.~Zahid}\ \bibnamefont
			{Hasan}},\ }\bibfield  {title} {\enquote {\bibinfo {title} {Momentum-space
				imaging of cooper pairing in a half-Dirac-gas topological superconductor},}\
	}\href {\doibase 10.1038/nphys3139} {\bibfield  {journal} {\bibinfo
		{journal} {Nat Phys}\ }\textbf {\bibinfo {volume} {10}},\ \bibinfo {pages}
	{943--950} (\bibinfo {year} {2014})}\BibitemShut {NoStop}%
\bibitem [{\citenamefont {Kobayashi}\ and\ \citenamefont
	{Sato}(2015)}]{Kobayashi15}%
\BibitemOpen
\bibfield  {author} {\bibinfo {author} {\bibfnamefont {Shingo}\ \bibnamefont
		{Kobayashi}}\ and\ \bibinfo {author} {\bibfnamefont {Masatoshi}\ \bibnamefont
		{Sato}},\ }\bibfield  {title} {\enquote {\bibinfo {title} {Topological
			superconductivity in Dirac semimetals},}\ }\href {\doibase
	10.1103/physrevlett.115.187001} {\bibfield  {journal} {\bibinfo  {journal}
		{Phys. Rev. Lett.}\ }\textbf {\bibinfo {volume} {115}},\ \bibinfo {pages}
	{187001} (\bibinfo {year} {2015})}\BibitemShut {NoStop}%
\bibitem [{\citenamefont {Alexander}\ \emph {et~al.}(2012)\citenamefont
	{Alexander}, \citenamefont {Chen}, \citenamefont {Matsumoto},\ and\
	\citenamefont {Kamien}}]{Alexander12}%
\BibitemOpen
\bibfield  {author} {\bibinfo {author} {\bibfnamefont {Gareth~P.}\
		\bibnamefont {Alexander}}, \bibinfo {author} {\bibfnamefont {Bryan Gin-ge}\
		\bibnamefont {Chen}}, \bibinfo {author} {\bibfnamefont {Elisabetta~A.}\
		\bibnamefont {Matsumoto}}, \ and\ \bibinfo {author} {\bibfnamefont
		{Randall~D.}\ \bibnamefont {Kamien}},\ }\bibfield  {title} {\enquote
	{\bibinfo {title} {\textit{Colloquium}: Disclination loops, point defects,
			and all that in nematic liquid crystals},}\ }\href {\doibase
	10.1103/RevModPhys.84.497} {\bibfield  {journal} {\bibinfo  {journal} {Rev.
			Mod. Phys.}\ }\textbf {\bibinfo {volume} {84}},\ \bibinfo {pages} {497--514}
	(\bibinfo {year} {2012})}\BibitemShut {NoStop}%
\bibitem [{\citenamefont {Kleman}\ and\ \citenamefont
	{Lavrentovich}(2006)}]{Kleman06}%
\BibitemOpen
\bibfield  {author} {\bibinfo {author} {\bibfnamefont {M.}~\bibnamefont
		{Kleman}}\ and\ \bibinfo {author} {\bibfnamefont {O.~D.}\ \bibnamefont
		{Lavrentovich}},\ }\bibfield  {title} {\enquote {\bibinfo {title}
		{Topological point defects in nematic liquid crystals},}\ }\href
{http://www.informaworld.com/10.1080/14786430600593016} {\bibfield  {journal}
	{\bibinfo  {journal} {Philosophical Magazine}\ }\textbf {\bibinfo {volume}
		{86}},\ \bibinfo {pages} {4117--4137} (\bibinfo {year} {2006})}\BibitemShut
{NoStop}%
\bibitem [{\citenamefont {Thiele}(1973)}]{Thiele73}%
\BibitemOpen
\bibfield  {author} {\bibinfo {author} {\bibfnamefont {A.~A.}\ \bibnamefont
		{Thiele}},\ }\bibfield  {title} {\enquote {\bibinfo {title} {Steady--state
			motion of magnetic of magnetic domains},}\ }\href
{http://link.aps.org/abstract/PRL/v30/p230} {\bibfield  {journal} {\bibinfo
		{journal} {Phys. Rev. Lett.}\ }\textbf {\bibinfo {volume} {30}},\ \bibinfo
	{pages} {230--233} (\bibinfo {year} {1973})}\BibitemShut {NoStop}%
\bibitem [{\citenamefont {Belavin}\ and\ \citenamefont
	{Polyakov}(1975)}]{Belavin75}%
\BibitemOpen
\bibfield  {author} {\bibinfo {author} {\bibfnamefont {A.~A.}\ \bibnamefont
		{Belavin}}\ and\ \bibinfo {author} {\bibfnamefont {A.~M.}\ \bibnamefont
		{Polyakov}},\ }\bibfield  {title} {\enquote {\bibinfo {title} {Metastable
			states of a {2D} isotropic ferromagnet},}\ }\href@noop {} {\bibfield
	{journal} {\bibinfo  {journal} {JETP Lett.}\ }\textbf {\bibinfo {volume}
		{22}},\ \bibinfo {pages} {245} (\bibinfo {year} {1975})}\BibitemShut
{NoStop}%
\bibitem [{\citenamefont {Malozemoff}\ and\ \citenamefont
	{Slonzewski}(1979)}]{Malozemoff79}%
\BibitemOpen
\bibfield  {author} {\bibinfo {author} {\bibfnamefont {A.~P.}\ \bibnamefont
		{Malozemoff}}\ and\ \bibinfo {author} {\bibfnamefont {J.~C.}\ \bibnamefont
		{Slonzewski}},\ }\href@noop {} {\emph {\bibinfo {title} {Magnetic domain
			walls in bubble materials}}}\ (\bibinfo  {publisher} {Academic Press},\
\bibinfo {address} {New York},\ \bibinfo {year} {1979})\BibitemShut {NoStop}%
\bibitem [{\citenamefont {Papanicolaou}\ and\ \citenamefont
	{Tomaras}(1991)}]{Papanicolaou91}%
\BibitemOpen
\bibfield  {author} {\bibinfo {author} {\bibfnamefont {N.}~\bibnamefont
		{Papanicolaou}}\ and\ \bibinfo {author} {\bibfnamefont {T.~N.}\ \bibnamefont
		{Tomaras}},\ }\bibfield  {title} {\enquote {\bibinfo {title} {Dynamics of
			magnetic vortices},}\ }\href
{http://www.sciencedirect.com/science/article/B6TVC-470F3HY-36/2/69a5e1a128fb5b7ef2ee0c512c3d78fc}
{\bibfield  {journal} {\bibinfo  {journal} {Nuclear Physics B}\ }\textbf
	{\bibinfo {volume} {360}},\ \bibinfo {pages} {425--462} (\bibinfo {year}
	{1991})}\BibitemShut {NoStop}%
\bibitem [{\citenamefont {Komineas}\ and\ \citenamefont
	{Papanicolaou}(1996)}]{Komineas96}%
\BibitemOpen
\bibfield  {author} {\bibinfo {author} {\bibfnamefont {S.}~\bibnamefont
		{Komineas}}\ and\ \bibinfo {author} {\bibfnamefont {N.}~\bibnamefont
		{Papanicolaou}},\ }\bibfield  {title} {\enquote {\bibinfo {title} {Topology
			and dynamics in ferromagnetic media},}\ }\href
{http://www.sciencedirect.com/science/article/B6TVK-3SPCRW6-6/2/f95e4629d5db97bbd9d8bdcbdbd476f3}
{\bibfield  {journal} {\bibinfo  {journal} {Physica D: Nonlinear Phenomena}\
	}\textbf {\bibinfo {volume} {99}},\ \bibinfo {pages} {81--107} (\bibinfo
	{year} {1996})}\BibitemShut {NoStop}%
\bibitem [{\citenamefont {Barker}\ and\ \citenamefont
	{Tretiakov}(2016)}]{Barker16}%
\BibitemOpen
\bibfield  {author} {\bibinfo {author} {\bibfnamefont {Joseph}\ \bibnamefont
		{Barker}}\ and\ \bibinfo {author} {\bibfnamefont {Oleg~A.}\ \bibnamefont
		{Tretiakov}},\ }\bibfield  {title} {\enquote {\bibinfo {title} {Static and
			dynamical properties of antiferromagnetic skyrmions in the presence of
			applied current and temperature},}\ }\href {\doibase
	10.1103/physrevlett.116.147203} {\bibfield  {journal} {\bibinfo  {journal}
		{Phys. Rev. Lett.}\ }\textbf {\bibinfo {volume} {116}},\ \bibinfo {pages}
	{147203} (\bibinfo {year} {2016})}\BibitemShut {NoStop}%
\bibitem [{\citenamefont {Bowick}\ and\ \citenamefont
	{Giomi}(2009)}]{Bowick09}%
\BibitemOpen
\bibfield  {author} {\bibinfo {author} {\bibfnamefont {Mark~J.}\ \bibnamefont
		{Bowick}}\ and\ \bibinfo {author} {\bibfnamefont {Luca}\ \bibnamefont
		{Giomi}},\ }\bibfield  {title} {\enquote {\bibinfo {title} {Two-dimensional
			matter: order, curvature and defects},}\ }\href {\doibase
	10.1080/00018730903043166} {\bibfield  {journal} {\bibinfo  {journal}
		{Advances in Physics}\ }\textbf {\bibinfo {volume} {58}},\ \bibinfo {pages}
	{449--563} (\bibinfo {year} {2009})}\BibitemShut {NoStop}%
\bibitem [{\citenamefont {Vitelli}\ and\ \citenamefont
	{Turner}(2004)}]{Vitelli04}%
\BibitemOpen
\bibfield  {author} {\bibinfo {author} {\bibfnamefont {Vincenzo}\
		\bibnamefont {Vitelli}}\ and\ \bibinfo {author} {\bibfnamefont {Ari~M.}\
		\bibnamefont {Turner}},\ }\bibfield  {title} {\enquote {\bibinfo {title}
		{Anomalous coupling between topological defects and curvature},}\ }\href
{\doibase 10.1103/PhysRevLett.93.215301} {\bibfield  {journal} {\bibinfo
		{journal} {Phys. Rev. Lett.}\ }\textbf {\bibinfo {volume} {93}},\ \bibinfo
	{pages} {215301} (\bibinfo {year} {2004})}\BibitemShut {NoStop}%
\bibitem [{\citenamefont {Turner}\ \emph {et~al.}(2010)\citenamefont {Turner},
	\citenamefont {Vitelli},\ and\ \citenamefont {Nelson}}]{Turner10}%
\BibitemOpen
\bibfield  {author} {\bibinfo {author} {\bibfnamefont {Ari~M.}\ \bibnamefont
		{Turner}}, \bibinfo {author} {\bibfnamefont {Vincenzo}\ \bibnamefont
		{Vitelli}}, \ and\ \bibinfo {author} {\bibfnamefont {David~R.}\ \bibnamefont
		{Nelson}},\ }\bibfield  {title} {\enquote {\bibinfo {title} {Vortices on
			curved surfaces},}\ }\href {\doibase 10.1103/RevModPhys.82.1301} {\bibfield
	{journal} {\bibinfo  {journal} {Rev. Mod. Phys.}\ }\textbf {\bibinfo {volume}
		{82}},\ \bibinfo {pages} {1301--1348} (\bibinfo {year} {2010})}\BibitemShut
{NoStop}%
\bibitem [{\citenamefont {Napoli}\ and\ \citenamefont
	{Vergori}(2012)}]{Napoli12}%
\BibitemOpen
\bibfield  {author} {\bibinfo {author} {\bibfnamefont {Gaetano}\ \bibnamefont
		{Napoli}}\ and\ \bibinfo {author} {\bibfnamefont {Luigi}\ \bibnamefont
		{Vergori}},\ }\bibfield  {title} {\enquote {\bibinfo {title} {Extrinsic
			curvature effects on nematic shells},}\ }\href {\doibase
	10.1103/PhysRevLett.108.207803} {\bibfield  {journal} {\bibinfo  {journal}
		{Physical Review Letters}\ }\textbf {\bibinfo {volume} {108}},\ \bibinfo
	{pages} {207803} (\bibinfo {year} {2012})}\BibitemShut {NoStop}%
\bibitem [{\citenamefont {Napoli}\ and\ \citenamefont
	{Vergori}(2013)}]{Napoli13}%
\BibitemOpen
\bibfield  {author} {\bibinfo {author} {\bibfnamefont {Gaetano}\ \bibnamefont
		{Napoli}}\ and\ \bibinfo {author} {\bibfnamefont {Luigi}\ \bibnamefont
		{Vergori}},\ }\bibfield  {title} {\enquote {\bibinfo {title} {Effective free
			energies for cholesteric shells},}\ }\href {\doibase 10.1039/c3sm50605c}
{\bibfield  {journal} {\bibinfo  {journal} {Soft Matter}\ }\textbf {\bibinfo
		{volume} {9}},\ \bibinfo {pages} {8378} (\bibinfo {year} {2013})}\BibitemShut
{NoStop}%
\bibitem [{\citenamefont {Gaididei}\ \emph {et~al.}(2014)\citenamefont
	{Gaididei}, \citenamefont {Kravchuk},\ and\ \citenamefont
	{Sheka}}]{Gaididei14}%
\BibitemOpen
\bibfield  {author} {\bibinfo {author} {\bibfnamefont {Yuri}\ \bibnamefont
		{Gaididei}}, \bibinfo {author} {\bibfnamefont {Volodymyr~P.}\ \bibnamefont
		{Kravchuk}}, \ and\ \bibinfo {author} {\bibfnamefont {Denis~D.}\ \bibnamefont
		{Sheka}},\ }\bibfield  {title} {\enquote {\bibinfo {title} {Curvature effects
			in thin magnetic shells},}\ }\href {\doibase 10.1103/PhysRevLett.112.257203}
{\bibfield  {journal} {\bibinfo  {journal} {Phys. Rev. Lett.}\ }\textbf
	{\bibinfo {volume} {112}},\ \bibinfo {pages} {257203} (\bibinfo {year}
	{2014})}\BibitemShut {NoStop}%
\bibitem [{\citenamefont {Sheka}\ \emph {et~al.}(2015)\citenamefont {Sheka},
	\citenamefont {Kravchuk},\ and\ \citenamefont {Gaididei}}]{Sheka15}%
\BibitemOpen
\bibfield  {author} {\bibinfo {author} {\bibfnamefont {Denis~D.}\
		\bibnamefont {Sheka}}, \bibinfo {author} {\bibfnamefont {Volodymyr~P.}\
		\bibnamefont {Kravchuk}}, \ and\ \bibinfo {author} {\bibfnamefont {Yuri}\
		\bibnamefont {Gaididei}},\ }\bibfield  {title} {\enquote {\bibinfo {title}
		{Curvature effects in statics and dynamics of low dimensional magnets},}\
}\href {\doibase http://dx.doi.org/10.1088/1751-8113/48/12/125202} {\bibfield
{journal} {\bibinfo  {journal} {Journal of Physics A: Mathematical and
		Theoretical}\ }\textbf {\bibinfo {volume} {48}},\ \bibinfo {pages} {125202}
(\bibinfo {year} {2015})}\BibitemShut {NoStop}%
\bibitem [{\citenamefont {Mermin}(1979)}]{Mermin79}%
\BibitemOpen
\bibfield  {author} {\bibinfo {author} {\bibfnamefont {N.~D.}\ \bibnamefont
		{Mermin}},\ }\bibfield  {title} {\enquote {\bibinfo {title} {The topological
			theory of defects in ordered media},}\ }\href {\doibase
	10.1103/revmodphys.51.591} {\bibfield  {journal} {\bibinfo  {journal}
		{Reviews of Modern Physics}\ }\textbf {\bibinfo {volume} {51}},\ \bibinfo
	{pages} {591--648} (\bibinfo {year} {1979})}\BibitemShut {NoStop}%
\bibitem [{\citenamefont {Thouless}(1998)}]{Thouless98}%
\BibitemOpen
\bibfield  {author} {\bibinfo {author} {\bibfnamefont {David~J.}\
		\bibnamefont {Thouless}},\ }\href@noop {} {\emph {\bibinfo {title}
		{Topological Quantum Numbers in Nonrelativistic Physics}}}\ (\bibinfo
{publisher} {World Scientific},\ \bibinfo {year} {1998})\BibitemShut
{NoStop}%
\bibitem [{\citenamefont {Dubrovin}\ \emph {et~al.}(1985)\citenamefont
	{Dubrovin}, \citenamefont {Fomenko},\ and\ \citenamefont
	{Novikov}}]{Dubrovin85p2}%
\BibitemOpen
\bibfield  {author} {\bibinfo {author} {\bibfnamefont {B.}~\bibnamefont
		{Dubrovin}}, \bibinfo {author} {\bibfnamefont {A.}~\bibnamefont {Fomenko}}, \
	and\ \bibinfo {author} {\bibfnamefont {S.}~\bibnamefont {Novikov}},\
}\href@noop {} {\emph {\bibinfo {title} {Modern Geometry - Methods and
		Applications: Part II: The Geometry and Topology of Manifolds}}},\ GTM093\
(\bibinfo  {publisher} {Springer},\ \bibinfo {year} {1985})\BibitemShut
{NoStop}%
\bibitem [{\citenamefont {Kosevich}\ \emph {et~al.}(1990)\citenamefont
	{Kosevich}, \citenamefont {Ivanov},\ and\ \citenamefont
	{Kovalev}}]{Kosevich90}%
\BibitemOpen
\bibfield  {author} {\bibinfo {author} {\bibfnamefont {A.~M.}\ \bibnamefont
		{Kosevich}}, \bibinfo {author} {\bibfnamefont {B.~A.}\ \bibnamefont
		{Ivanov}}, \ and\ \bibinfo {author} {\bibfnamefont {A.~S.}\ \bibnamefont
		{Kovalev}},\ }\bibfield  {title} {\enquote {\bibinfo {title} {Magnetic
			solitons},}\ }\href
{http://www.sciencedirect.com/science/article/B6TVP-46SXP0P-19/2/d0a99bbbccf078602123db3e5d601202}
{\bibfield  {journal} {\bibinfo  {journal} {Physics Reports}\ }\textbf
	{\bibinfo {volume} {194}},\ \bibinfo {pages} {117--238} (\bibinfo {year}
	{1990})}\BibitemShut {NoStop}%
\bibitem [{\citenamefont {Manton}\ and\ \citenamefont
	{Sutcliffe}(2004)}]{Manton04}%
\BibitemOpen
\bibfield  {author} {\bibinfo {author} {\bibfnamefont {Nicholas}\
		\bibnamefont {Manton}}\ and\ \bibinfo {author} {\bibfnamefont {Pau}\
		\bibnamefont {Sutcliffe}},\ }\href@noop {} {\emph {\bibinfo {title}
		{Topological solitons}}},\ Cambridge Monographs on Mathematical Physics\
(\bibinfo  {publisher} {Cambridge University Press},\ \bibinfo {address}
{Cambridge},\ \bibinfo {year} {2004})\BibitemShut {NoStop}%
\bibitem [{\citenamefont {Bogdanov}\ and\ \citenamefont
	{Yablonski\u{\i}}(1989)}]{Bogdanov89}%
\BibitemOpen
\bibfield  {author} {\bibinfo {author} {\bibfnamefont {A.~N.}\ \bibnamefont
		{Bogdanov}}\ and\ \bibinfo {author} {\bibfnamefont {D.~A.}\ \bibnamefont
		{Yablonski\u{\i}}},\ }\bibfield  {title} {\enquote {\bibinfo {title}
		{Thermodynamically stable `` vortices'' in magnetically ordered crystals. the
			mixed state of magnets},}\ }\href@noop {} {\bibfield  {journal} {\bibinfo
		{journal} {Zh. Eksp. Teor. Fiz.}\ }\textbf {\bibinfo {volume} {95}},\
	\bibinfo {pages} {178--182} (\bibinfo {year} {1989})}\BibitemShut {NoStop}%
\bibitem [{\citenamefont {Bogdanov}\ and\ \citenamefont
	{Hubert}(1994{\natexlab{a}})}]{Bogdanov94}%
\BibitemOpen
\bibfield  {author} {\bibinfo {author} {\bibfnamefont {A.}~\bibnamefont
		{Bogdanov}}\ and\ \bibinfo {author} {\bibfnamefont {A.}~\bibnamefont
		{Hubert}},\ }\bibfield  {title} {\enquote {\bibinfo {title}
		{Thermodynamically stable magnetic vortex states in magnetic crystals},}\
}\href {\doibase 10.1016/0304-8853(94)90046-9} {\bibfield  {journal}
{\bibinfo  {journal} {Journal of Magnetism and Magnetic Materials}\ }\textbf
{\bibinfo {volume} {138}},\ \bibinfo {pages} {255--269} (\bibinfo {year}
{1994}{\natexlab{a}})}\BibitemShut {NoStop}%
\bibitem [{\citenamefont {Bogdanov}\ and\ \citenamefont
	{Hubert}(1999)}]{Bogdanov99}%
\BibitemOpen
\bibfield  {author} {\bibinfo {author} {\bibfnamefont {A.}~\bibnamefont
		{Bogdanov}}\ and\ \bibinfo {author} {\bibfnamefont {A.}~\bibnamefont
		{Hubert}},\ }\bibfield  {title} {\enquote {\bibinfo {title} {The stability of
			vortex-like structures in uniaxial ferromagnets},}\ }\href {\doibase
	10.1016/s0304-8853(98)01038-5} {\bibfield  {journal} {\bibinfo  {journal}
		{Journal of Magnetism and Magnetic Materials}\ }\textbf {\bibinfo {volume}
		{195}},\ \bibinfo {pages} {182--192} (\bibinfo {year} {1999})}\BibitemShut
{NoStop}%
\bibitem [{\citenamefont {Bogdanov}\ and\ \citenamefont
	{R{\"o}{\ss}ler}(2001)}]{Bogdanov01}%
\BibitemOpen
\bibfield  {author} {\bibinfo {author} {\bibfnamefont {A.}~\bibnamefont
		{Bogdanov}}\ and\ \bibinfo {author} {\bibfnamefont {U.}~\bibnamefont
		{R{\"o}{\ss}ler}},\ }\bibfield  {title} {\enquote {\bibinfo {title} {Chiral
			symmetry breaking in magnetic thin films and multilayers},}\ }\href {\doibase
	10.1103/physrevlett.87.037203} {\bibfield  {journal} {\bibinfo  {journal}
		{Physical Review Letters}\ }\textbf {\bibinfo {volume} {87}},\ \bibinfo
	{pages} {037203} (\bibinfo {year} {2001})}\BibitemShut {NoStop}%
\bibitem [{\citenamefont {Romming}\ \emph {et~al.}(2013)\citenamefont
	{Romming}, \citenamefont {Hanneken}, \citenamefont {Menzel}, \citenamefont
	{Bickel}, \citenamefont {Wolter}, \citenamefont {von Bergmann}, \citenamefont
	{Kubetzka},\ and\ \citenamefont {Wiesendanger}}]{Romming13}%
\BibitemOpen
\bibfield  {author} {\bibinfo {author} {\bibfnamefont {Niklas}\ \bibnamefont
		{Romming}}, \bibinfo {author} {\bibfnamefont {Christian}\ \bibnamefont
		{Hanneken}}, \bibinfo {author} {\bibfnamefont {Matthias}\ \bibnamefont
		{Menzel}}, \bibinfo {author} {\bibfnamefont {Jessica~E.}\ \bibnamefont
		{Bickel}}, \bibinfo {author} {\bibfnamefont {Boris}\ \bibnamefont {Wolter}},
	\bibinfo {author} {\bibfnamefont {Kirsten}\ \bibnamefont {von Bergmann}},
	\bibinfo {author} {\bibfnamefont {AndrГ©}\ \bibnamefont {Kubetzka}}, \ and\
	\bibinfo {author} {\bibfnamefont {Roland}\ \bibnamefont {Wiesendanger}},\
}\bibfield  {title} {\enquote {\bibinfo {title} {Writing and deleting single
		magnetic skyrmions},}\ }\href {\doibase 10.1126/science.1240573} {\bibfield
{journal} {\bibinfo  {journal} {Science}\ }\textbf {\bibinfo {volume}
	{341}},\ \bibinfo {pages} {636--639} (\bibinfo {year} {2013})}\  \BibitemShut
{NoStop}%
\bibitem [{\citenamefont {Romming}\ \emph {et~al.}(2015)\citenamefont
	{Romming}, \citenamefont {Kubetzka}, \citenamefont {Hanneken}, \citenamefont
	{von Bergmann},\ and\ \citenamefont {Wiesendanger}}]{Romming15}%
\BibitemOpen
\bibfield  {author} {\bibinfo {author} {\bibfnamefont {Niklas}\ \bibnamefont
		{Romming}}, \bibinfo {author} {\bibfnamefont {Andr{\'{e}}}\ \bibnamefont
		{Kubetzka}}, \bibinfo {author} {\bibfnamefont {Christian}\ \bibnamefont
		{Hanneken}}, \bibinfo {author} {\bibfnamefont {Kirsten}\ \bibnamefont {von
			Bergmann}}, \ and\ \bibinfo {author} {\bibfnamefont {Roland}\ \bibnamefont
		{Wiesendanger}},\ }\bibfield  {title} {\enquote {\bibinfo {title}
		{Field-dependent size and shape of single magnetic skyrmions},}\ }\href
{\doibase 10.1103/physrevlett.114.177203} {\bibfield  {journal} {\bibinfo
		{journal} {Phys. Rev. Lett.}\ }\textbf {\bibinfo {volume} {114}},\ \bibinfo
	{pages} {177203} (\bibinfo {year} {2015})}\BibitemShut {NoStop}%
\bibitem [{\citenamefont {B{\"u}ttner}\ \emph {et~al.}(2015)\citenamefont
	{B{\"u}ttner}, \citenamefont {Moutafis}, \citenamefont {Schneider},
	\citenamefont {Kr{\"u}ger}, \citenamefont {G{\"u}nther}, \citenamefont
	{Geilhufe}, \citenamefont {v.~Korff~Schmising}, \citenamefont {Mohanty},
	\citenamefont {Pfau}, \citenamefont {Schaffert}, \citenamefont {Bisig},
	\citenamefont {Foerster}, \citenamefont {Schulz}, \citenamefont {Vaz},
	\citenamefont {Franken}, \citenamefont {Swagten}, \citenamefont {Kl{\"a}ui},\
	and\ \citenamefont {Eisebitt}}]{Buettner15}%
\BibitemOpen
\bibfield  {author} {\bibinfo {author} {\bibfnamefont {Felix}\ \bibnamefont
		{B{\"u}ttner}}, \bibinfo {author} {\bibfnamefont {C.}~\bibnamefont
		{Moutafis}}, \bibinfo {author} {\bibfnamefont {M.}~\bibnamefont {Schneider}},
	\bibinfo {author} {\bibfnamefont {B.}~\bibnamefont {Kr{\"u}ger}}, \bibinfo
	{author} {\bibfnamefont {C.~M.}\ \bibnamefont {G{\"u}nther}}, \bibinfo
	{author} {\bibfnamefont {J.}~\bibnamefont {Geilhufe}}, \bibinfo {author}
	{\bibfnamefont {C.}~\bibnamefont {v.~Korff~Schmising}}, \bibinfo {author}
	{\bibfnamefont {J.}~\bibnamefont {Mohanty}}, \bibinfo {author} {\bibfnamefont
		{B.}~\bibnamefont {Pfau}}, \bibinfo {author} {\bibfnamefont {S.}~\bibnamefont
		{Schaffert}}, \bibinfo {author} {\bibfnamefont {A.}~\bibnamefont {Bisig}},
	\bibinfo {author} {\bibfnamefont {M.}~\bibnamefont {Foerster}}, \bibinfo
	{author} {\bibfnamefont {T.}~\bibnamefont {Schulz}}, \bibinfo {author}
	{\bibfnamefont {C.~A.~F.}\ \bibnamefont {Vaz}}, \bibinfo {author}
	{\bibfnamefont {J.~H.}\ \bibnamefont {Franken}}, \bibinfo {author}
	{\bibfnamefont {H.~J.~M.}\ \bibnamefont {Swagten}}, \bibinfo {author}
	{\bibfnamefont {M.}~\bibnamefont {Kl{\"a}ui}}, \ and\ \bibinfo {author}
	{\bibfnamefont {S.}~\bibnamefont {Eisebitt}},\ }\bibfield  {title} {\enquote
	{\bibinfo {title} {Dynamics and inertia of skyrmionic spin structures},}\
}\href {\doibase 10.1038/nphys3234} {\bibfield  {journal} {\bibinfo
	{journal} {Nat Phys}\ }\textbf {\bibinfo {volume} {11}},\ \bibinfo {pages}
{225--228} (\bibinfo {year} {2015})}\BibitemShut {NoStop}%
\bibitem [{\citenamefont {Leonov}\ \emph {et~al.}(2016)\citenamefont {Leonov},
	\citenamefont {Monchesky}, \citenamefont {Romming}, \citenamefont {Kubetzka},
	\citenamefont {Bogdanov},\ and\ \citenamefont {Wiesendanger}}]{Leonov16}%
\BibitemOpen
\bibfield  {author} {\bibinfo {author} {\bibfnamefont {A~O}\ \bibnamefont
		{Leonov}}, \bibinfo {author} {\bibfnamefont {T~L}\ \bibnamefont {Monchesky}},
	\bibinfo {author} {\bibfnamefont {N}~\bibnamefont {Romming}}, \bibinfo
	{author} {\bibfnamefont {A}~\bibnamefont {Kubetzka}}, \bibinfo {author}
	{\bibfnamefont {A~N}\ \bibnamefont {Bogdanov}}, \ and\ \bibinfo {author}
	{\bibfnamefont {R}~\bibnamefont {Wiesendanger}},\ }\bibfield  {title}
{\enquote {\bibinfo {title} {The properties of isolated chiral skyrmions in
			thin magnetic films},}\ }\href {\doibase 10.1088/1367-2630/18/6/065003}
{\bibfield  {journal} {\bibinfo  {journal} {New J. Phys.}\ }\textbf {\bibinfo
		{volume} {18}},\ \bibinfo {pages} {065003} (\bibinfo {year}
	{2016})}\BibitemShut {NoStop}%
\bibitem [{\citenamefont {Komineas}\ and\ \citenamefont
	{Papanicolaou}(2015)}]{Komineas15c}%
\BibitemOpen
\bibfield  {author} {\bibinfo {author} {\bibfnamefont {Stavros}\ \bibnamefont
		{Komineas}}\ and\ \bibinfo {author} {\bibfnamefont {Nikos}\ \bibnamefont
		{Papanicolaou}},\ }\bibfield  {title} {\enquote {\bibinfo {title} {Skyrmion
			dynamics in chiral ferromagnets},}\ }\href {\doibase
	10.1103/physrevb.92.064412} {\bibfield  {journal} {\bibinfo  {journal} {Phys.
			Rev. B}\ }\textbf {\bibinfo {volume} {92}} (\bibinfo {year} {2015})}\BibitemShut {NoStop}%
\bibitem [{\citenamefont {Machon}\ and\ \citenamefont
	{Alexander}(2013)}]{Machon13}%
\BibitemOpen
\bibfield  {author} {\bibinfo {author} {\bibfnamefont {Thomas}\ \bibnamefont
		{Machon}}\ and\ \bibinfo {author} {\bibfnamefont {Gareth~P.}\ \bibnamefont
		{Alexander}},\ }\bibfield  {title} {\enquote {\bibinfo {title} {Knots and
			nonorientable surfaces in chiral nematics},}\ }\href {\doibase
	10.1073/pnas.1308225110} {\bibfield  {journal} {\bibinfo  {journal}
		{Proceedings of the National Academy of Sciences}\ }\textbf {\bibinfo
		{volume} {110}},\ \bibinfo {pages} {14174--14179} (\bibinfo {year}
	{2013})}\BibitemShut {NoStop}%
\bibitem [{\citenamefont {Jampani}\ \emph {et~al.}(2011)\citenamefont
	{Jampani}, \citenamefont {Skarabot}, \citenamefont {Ravnik}, \citenamefont
	{Copar}, \citenamefont {Zumer},\ and\ \citenamefont {Musevic}}]{Jampani11}%
\BibitemOpen
\bibfield  {author} {\bibinfo {author} {\bibfnamefont {V.~S.~R.}\
		\bibnamefont {Jampani}}, \bibinfo {author} {\bibfnamefont {M.}~\bibnamefont
		{Skarabot}}, \bibinfo {author} {\bibfnamefont {M.}~\bibnamefont {Ravnik}},
	\bibinfo {author} {\bibfnamefont {S.}~\bibnamefont {Copar}}, \bibinfo
	{author} {\bibfnamefont {S.}~\bibnamefont {Zumer}}, \ and\ \bibinfo {author}
	{\bibfnamefont {I.}~\bibnamefont {Musevic}},\ }\bibfield  {title} {\enquote
	{\bibinfo {title} {Colloidal entanglement in highly twisted chiral nematic
			colloids: Twisted loops, hopf links, and trefoil knots},}\ }\href {\doibase
	10.1103/physreve.84.031703} {\bibfield  {journal} {\bibinfo  {journal} {Phys.
			Rev. E}\ }\textbf {\bibinfo {volume} {84}},\ \bibinfo {pages} {031703}
	(\bibinfo {year} {2011})}\BibitemShut {NoStop}%
\bibitem [{\citenamefont {Senyuk}\ \emph {et~al.}(2012)\citenamefont {Senyuk},
	\citenamefont {Liu}, \citenamefont {He}, \citenamefont {Kamien},
	\citenamefont {Kusner}, \citenamefont {Lubensky},\ and\ \citenamefont
	{Smalyukh}}]{Senyuk12}%
\BibitemOpen
\bibfield  {author} {\bibinfo {author} {\bibfnamefont {Bohdan}\ \bibnamefont
		{Senyuk}}, \bibinfo {author} {\bibfnamefont {Qingkun}\ \bibnamefont {Liu}},
	\bibinfo {author} {\bibfnamefont {Sailing}\ \bibnamefont {He}}, \bibinfo
	{author} {\bibfnamefont {Randall~D.}\ \bibnamefont {Kamien}}, \bibinfo
	{author} {\bibfnamefont {Robert~B.}\ \bibnamefont {Kusner}}, \bibinfo
	{author} {\bibfnamefont {Tom~C.}\ \bibnamefont {Lubensky}}, \ and\ \bibinfo
	{author} {\bibfnamefont {Ivan~I.}\ \bibnamefont {Smalyukh}},\ }\bibfield
{title} {\enquote {\bibinfo {title} {Topological colloids},}\ }\href
{\doibase 10.1038/nature11710} {\bibfield  {journal} {\bibinfo  {journal}
		{Nature}\ }\textbf {\bibinfo {volume} {493}},\ \bibinfo {pages} {200--205}
	(\bibinfo {year} {2012})}\BibitemShut {NoStop}%
\bibitem [{\citenamefont {Mertens}\ and\ \citenamefont
	{Bishop}(2000)}]{Mertens00}%
\BibitemOpen
\bibfield  {author} {\bibinfo {author} {\bibfnamefont {F.~G.}\ \bibnamefont
		{Mertens}}\ and\ \bibinfo {author} {\bibfnamefont {A.~R.}\ \bibnamefont
		{Bishop}},\ }\bibfield  {title} {\enquote {\bibinfo {title} {Dynamics of
			vortices in two--dimensional magnets},}\ }in\ \href@noop {} {\emph {\bibinfo
		{booktitle} {Nonlinear Science at the Dawn of the 21th Century}}},\ \bibinfo
{editor} {edited by\ \bibinfo {editor} {\bibfnamefont {P.~L.}\ \bibnamefont
		{Christiansen}}, \bibinfo {editor} {\bibfnamefont {M.~P.}\ \bibnamefont
		{Soerensen}}, \ and\ \bibinfo {editor} {\bibfnamefont {A.~C.}\ \bibnamefont
		{Scott}}}\ (\bibinfo  {publisher} {Springer--Verlag},\ \bibinfo {address}
{Berlin},\ \bibinfo {year} {2000})\ pp.\ \bibinfo {pages}
{137--170}\BibitemShut {NoStop}%
\bibitem [{\citenamefont {Feldtkeller}(1965)}]{Feldtkeller65b}%
\BibitemOpen
\bibfield  {author} {\bibinfo {author} {\bibfnamefont {Ernst}\ \bibnamefont
		{Feldtkeller}},\ }\bibfield  {title} {\enquote {\bibinfo {title}
		{Mikromagnetisch stetige und unstetige magnetisierungskonfigurationen},}\
}\href@noop {} {\bibfield  {journal} {\bibinfo  {journal} {Zeitschrift
		f\"{u}r angewandte Physik}\ }\textbf {\bibinfo {volume} {19}},\ \bibinfo
{pages} {530--536} (\bibinfo {year} {1965})}\BibitemShut {NoStop}%
\bibitem [{\citenamefont {Sampaio}\ \emph {et~al.}(2013)\citenamefont
	{Sampaio}, \citenamefont {Cros}, \citenamefont {Rohart}, \citenamefont
	{Thiaville},\ and\ \citenamefont {Fert}}]{Sampaio13}%
\BibitemOpen
\bibfield  {author} {\bibinfo {author} {\bibfnamefont {J.}~\bibnamefont
		{Sampaio}}, \bibinfo {author} {\bibfnamefont {V.}~\bibnamefont {Cros}},
	\bibinfo {author} {\bibfnamefont {S.}~\bibnamefont {Rohart}}, \bibinfo
	{author} {\bibfnamefont {A.}~\bibnamefont {Thiaville}}, \ and\ \bibinfo
	{author} {\bibfnamefont {A.}~\bibnamefont {Fert}},\ }\bibfield  {title}
{\enquote {\bibinfo {title} {Nucleation, stability and current-induced motion
			of isolated magnetic skyrmions in nanostructures},}\ }\href {\doibase
	10.1038/nnano.2013.210} {\bibfield  {journal} {\bibinfo  {journal} {Nature
			Nanotechnology}\ }\textbf {\bibinfo {volume} {8}},\ \bibinfo {pages}
	{839--844} (\bibinfo {year} {2013})}\BibitemShut {NoStop}%
\bibitem [{\citenamefont {Rohart}\ and\ \citenamefont
	{Thiaville}(2013)}]{Rohart13}%
\BibitemOpen
\bibfield  {author} {\bibinfo {author} {\bibfnamefont {S.}~\bibnamefont
		{Rohart}}\ and\ \bibinfo {author} {\bibfnamefont {A.}~\bibnamefont
		{Thiaville}},\ }\bibfield  {title} {\enquote {\bibinfo {title} {Skyrmion
			confinement in ultrathin film nanostructures in the presence of
			Dzyaloshinskii-Moriya interaction},}\ }\href {\doibase
	10.1103/PhysRevB.88.184422} {\bibfield  {journal} {\bibinfo  {journal}
		{Physical Review B}\ }\textbf {\bibinfo {volume} {88}},\ \bibinfo {pages}
	{184422} (\bibinfo {year} {2013})}\BibitemShut {NoStop}%
\bibitem [{\citenamefont {Rothman}\ \emph {et~al.}(2001)\citenamefont
	{Rothman}, \citenamefont {Kl{\"a}ui}, \citenamefont {Lopez-Diaz},
	\citenamefont {Vaz}, \citenamefont {Bleloch}, \citenamefont {Bland},
	\citenamefont {Cui},\ and\ \citenamefont {Speaks}}]{Rothman01}%
\BibitemOpen
\bibfield  {author} {\bibinfo {author} {\bibfnamefont {J.}~\bibnamefont
		{Rothman}}, \bibinfo {author} {\bibfnamefont {M.}~\bibnamefont {Kl{\"a}ui}},
	\bibinfo {author} {\bibfnamefont {L.}~\bibnamefont {Lopez-Diaz}}, \bibinfo
	{author} {\bibfnamefont {C.~A.~F.}\ \bibnamefont {Vaz}}, \bibinfo {author}
	{\bibfnamefont {A.}~\bibnamefont {Bleloch}}, \bibinfo {author} {\bibfnamefont
		{J.~A.~C.}\ \bibnamefont {Bland}}, \bibinfo {author} {\bibfnamefont
		{Z.}~\bibnamefont {Cui}}, \ and\ \bibinfo {author} {\bibfnamefont
		{R.}~\bibnamefont {Speaks}},\ }\bibfield  {title} {\enquote {\bibinfo {title}
		{Observation of a bi-domain state and nucleation free switching in mesoscopic
			ring magnets},}\ }\href {\doibase 10.1103/physrevlett.86.1098} {\bibfield
	{journal} {\bibinfo  {journal} {Phys. Rev. Lett.}\ }\textbf {\bibinfo
		{volume} {86}},\ \bibinfo {pages} {1098--1101} (\bibinfo {year}
	{2001})}\BibitemShut {NoStop}%
\bibitem [{\citenamefont {Kamien}(2002)}]{Kamien02}%
\BibitemOpen
\bibfield  {author} {\bibinfo {author} {\bibfnamefont {Randall}\ \bibnamefont
		{Kamien}},\ }\bibfield  {title} {\enquote {\bibinfo {title} {The geometry of
			soft materials: a primer},}\ }\href {\doibase 10.1103/RevModPhys.74.953}
{\bibfield  {journal} {\bibinfo  {journal} {Reviews of Modern Physics}\
	}\textbf {\bibinfo {volume} {74}},\ \bibinfo {pages} {953--971} (\bibinfo
	{year} {2002})}\BibitemShut {NoStop}%
\bibitem [{Note1()}]{Note1}%
\BibitemOpen
\bibinfo {note} {For a planar film $Q_\protect \textsc {g}=0$.}\BibitemShut
{Stop}%
\bibitem [{\citenamefont {Landeros}\ and\ \citenamefont
	{N\'{u}{\~n}ez}(2010)}]{Landeros10}%
\BibitemOpen
\bibfield  {author} {\bibinfo {author} {\bibfnamefont {P.}~\bibnamefont
		{Landeros}}\ and\ \bibinfo {author} {\bibfnamefont {\'{A}lvaro~S.}\
		\bibnamefont {N\'{u}{\~n}ez}},\ }\bibfield  {title} {\enquote {\bibinfo
		{title} {Domain wall motion on magnetic nanotubes},}\ }\href {\doibase
	10.1063/1.3466747} {\bibfield  {journal} {\bibinfo  {journal} {Journal of
			Applied Physics}\ }\textbf {\bibinfo {volume} {108}},\ \bibinfo {eid}
	{033917} (\bibinfo {year} {2010})}\BibitemShut {NoStop}%
\bibitem [{\citenamefont {Yan}\ \emph {et~al.}(2011)\citenamefont {Yan},
	\citenamefont {Andreas}, \citenamefont {K{\'a}kay}, \citenamefont
	{Garcia-Sanchez},\ and\ \citenamefont {Hertel}}]{Yan11a}%
\BibitemOpen
\bibfield  {author} {\bibinfo {author} {\bibfnamefont {Ming}\ \bibnamefont
		{Yan}}, \bibinfo {author} {\bibfnamefont {Christian}\ \bibnamefont
		{Andreas}}, \bibinfo {author} {\bibfnamefont {Attila}\ \bibnamefont
		{K{\'a}kay}}, \bibinfo {author} {\bibfnamefont {Felipe}\ \bibnamefont
		{Garcia-Sanchez}}, \ and\ \bibinfo {author} {\bibfnamefont {Riccardo}\
		\bibnamefont {Hertel}},\ }\bibfield  {title} {\enquote {\bibinfo {title}
		{Fast domain wall dynamics in magnetic nanotubes: Suppression of walker
			breakdown and Cherenkov-like spin wave emission},}\ }\href {\doibase
	10.1063/1.3643037} {\bibfield  {journal} {\bibinfo  {journal} {Applied
			Physics Letters}\ }\textbf {\bibinfo {volume} {99}},\ \bibinfo {eid} {122505}
	(\bibinfo {year} {2011})}\BibitemShut {NoStop}%
\bibitem [{\citenamefont {Villain-Guillot}\ \emph {et~al.}(1995)\citenamefont
	{Villain-Guillot}, \citenamefont {Dandoloff}, \citenamefont {Saxena},\ and\
	\citenamefont {Bishop}}]{Villain-Guillot95}%
\BibitemOpen
\bibfield  {author} {\bibinfo {author} {\bibfnamefont {S.}~\bibnamefont
		{Villain-Guillot}}, \bibinfo {author} {\bibfnamefont {R.}~\bibnamefont
		{Dandoloff}}, \bibinfo {author} {\bibfnamefont {A.}~\bibnamefont {Saxena}}, \
	and\ \bibinfo {author} {\bibfnamefont {A.~R.}\ \bibnamefont {Bishop}},\
}\bibfield  {title} {\enquote {\bibinfo {title} {Topological solitons and
		geometrical frustration},}\ }\href {\doibase 10.1103/PhysRevB.52.6712}
{\bibfield  {journal} {\bibinfo  {journal} {Phys. Rev. B}\ }\textbf {\bibinfo
		{volume} {52}},\ \bibinfo {pages} {6712--6722} (\bibinfo {year}
	{1995})}\BibitemShut {NoStop}%
\bibitem [{\citenamefont {Milagre}\ and\ \citenamefont
	{Moura-Melo}(2007)}]{Milagre07}%
\BibitemOpen
\bibfield  {author} {\bibinfo {author} {\bibfnamefont {G.S.}\ \bibnamefont
		{Milagre}}\ and\ \bibinfo {author} {\bibfnamefont {Winder~A.}\ \bibnamefont
		{Moura-Melo}},\ }\bibfield  {title} {\enquote {\bibinfo {title} {Magnetic
			vortex-like excitations on a sphere},}\ }\href
{http://www.sciencedirect.com/science/article/B6TVM-4NDMN8T-6/1/5884027a6e920c09249e2e818d0d3e71}
{\bibfield  {journal} {\bibinfo  {journal} {Physics Letters A}\ }\textbf
	{\bibinfo {volume} {368}},\ \bibinfo {pages} {155--163} (\bibinfo {year}
	{2007})}\BibitemShut {NoStop}%
\bibitem [{\citenamefont {Kravchuk}\ \emph {et~al.}(2012)\citenamefont
	{Kravchuk}, \citenamefont {Sheka}, \citenamefont {Streubel}, \citenamefont
	{Makarov}, \citenamefont {Schmidt},\ and\ \citenamefont
	{Gaididei}}]{Kravchuk12a}%
\BibitemOpen
\bibfield  {author} {\bibinfo {author} {\bibfnamefont {Volodymyr~P.}\
		\bibnamefont {Kravchuk}}, \bibinfo {author} {\bibfnamefont {Denis~D.}\
		\bibnamefont {Sheka}}, \bibinfo {author} {\bibfnamefont {Robert}\
		\bibnamefont {Streubel}}, \bibinfo {author} {\bibfnamefont {Denys}\
		\bibnamefont {Makarov}}, \bibinfo {author} {\bibfnamefont {Oliver~G.}\
		\bibnamefont {Schmidt}}, \ and\ \bibinfo {author} {\bibfnamefont {Yuri}\
		\bibnamefont {Gaididei}},\ }\bibfield  {title} {\enquote {\bibinfo {title}
		{Out-of-surface vortices in spherical shells},}\ }\href {\doibase
	10.1103/PhysRevB.85.144433} {\bibfield  {journal} {\bibinfo  {journal} {Phys.
			Rev. B}\ }\textbf {\bibinfo {volume} {85}},\ \bibinfo {pages} {144433}
	(\bibinfo {year} {2012})}\BibitemShut {NoStop}%
\bibitem [{\citenamefont {Komineas}(2007)}]{Komineas07a}%
\BibitemOpen
\bibfield  {author} {\bibinfo {author} {\bibfnamefont {S.}~\bibnamefont
		{Komineas}},\ }\bibfield  {title} {\enquote {\bibinfo {title} {Rotating
			vortex dipoles in ferromagnets},}\ }\href {\doibase
	10.1103/PhysRevLett.99.117202} {\bibfield  {journal} {\bibinfo  {journal}
		{Phys. Rev. Lett.}\ }\textbf {\bibinfo {volume} {99}},\ \bibinfo {eid}
	{117202} (\bibinfo {year} {2007})}\BibitemShut {NoStop}%
\bibitem [{Note2()}]{Note2}%
\BibitemOpen
\bibinfo {note} {In fact, seminal works on magnetic skyrmions,\cite
	{Bogdanov89,Bogdanov94,Bogdanov99,Bogdanov01} used the term ``vortex''
	instead of ``skyrmion''.}\BibitemShut {Stop}%
\bibitem [{\citenamefont {Huber}(1982)}]{Huber82a}%
\BibitemOpen
\bibfield  {author} {\bibinfo {author} {\bibfnamefont {D.~L.}\ \bibnamefont
		{Huber}},\ }\bibfield  {title} {\enquote {\bibinfo {title} {Equation of
			motion of a spin vortex in a two-dimensional planar magnet},}\ }\href
{\doibase 10.1063/1.330661} {\bibfield  {journal} {\bibinfo  {journal}
		{J.~Appl. Phys.}\ }\textbf {\bibinfo {volume} {53}},\ \bibinfo {pages}
	{1899--1900} (\bibinfo {year} {1982})}\BibitemShut {NoStop}%
\bibitem [{\citenamefont {Lin}\ \emph {et~al.}(2013{\natexlab{a}})\citenamefont
	{Lin}, \citenamefont {Reichhardt}, \citenamefont {Batista},\ and\
	\citenamefont {Saxena}}]{Lin13}%
\BibitemOpen
\bibfield  {author} {\bibinfo {author} {\bibfnamefont {Shi-Zeng}\
		\bibnamefont {Lin}}, \bibinfo {author} {\bibfnamefont {Charles}\ \bibnamefont
		{Reichhardt}}, \bibinfo {author} {\bibfnamefont {Cristian~D.}\ \bibnamefont
		{Batista}}, \ and\ \bibinfo {author} {\bibfnamefont {Avadh}\ \bibnamefont
		{Saxena}},\ }\bibfield  {title} {\enquote {\bibinfo {title} {Driven skyrmions
			and dynamical transitions in chiral magnets},}\ }\href {\doibase
	10.1103/physrevlett.110.207202} {\bibfield  {journal} {\bibinfo  {journal}
		{Phys. Rev. Lett.}\ }\textbf {\bibinfo {volume} {110}},\ \bibinfo {pages}
	{207202} (\bibinfo {year} {2013}{\natexlab{a}})}\BibitemShut {NoStop}%
\bibitem [{\citenamefont {Lin}\ \emph {et~al.}(2013{\natexlab{b}})\citenamefont
	{Lin}, \citenamefont {Reichhardt}, \citenamefont {Batista},\ and\
	\citenamefont {Saxena}}]{Lin13b}%
\BibitemOpen
\bibfield  {author} {\bibinfo {author} {\bibfnamefont {Shi-Zeng}\
		\bibnamefont {Lin}}, \bibinfo {author} {\bibfnamefont {Charles}\ \bibnamefont
		{Reichhardt}}, \bibinfo {author} {\bibfnamefont {Cristian~D.}\ \bibnamefont
		{Batista}}, \ and\ \bibinfo {author} {\bibfnamefont {Avadh}\ \bibnamefont
		{Saxena}},\ }\bibfield  {title} {\enquote {\bibinfo {title} {Particle model
			for skyrmions in metallic chiral magnets: Dynamics, pinning, and creep},}\
}\href {\doibase 10.1103/physrevb.87.214419} {\bibfield  {journal} {\bibinfo
	{journal} {Phys. Rev. B}\ }\textbf {\bibinfo {volume} {87}},\ \bibinfo
{pages} {214419} (\bibinfo {year} {2013}{\natexlab{b}})}\BibitemShut
{NoStop}%
\bibitem [{\citenamefont {Lin}\ and\ \citenamefont {Saxena}(2015)}]{Lin15a}%
\BibitemOpen
\bibfield  {author} {\bibinfo {author} {\bibfnamefont {Shi-Zeng}\
		\bibnamefont {Lin}}\ and\ \bibinfo {author} {\bibfnamefont {Avadh}\
		\bibnamefont {Saxena}},\ }\bibfield  {title} {\enquote {\bibinfo {title}
		{Noncircular skyrmion and its anisotropic response in thin films of chiral
			magnets under a tilted magnetic field},}\ }\href {\doibase
	10.1103/physrevb.92.180401} {\bibfield  {journal} {\bibinfo  {journal} {Phys.
			Rev. B}\ }\textbf {\bibinfo {volume} {92}},\ \bibinfo {pages} {180401}
	(\bibinfo {year} {2015})}\BibitemShut {NoStop}%
\bibitem [{\citenamefont {Milde}\ \emph {et~al.}(2013)\citenamefont {Milde},
	\citenamefont {Kohler}, \citenamefont {Seidel}, \citenamefont {Eng},
	\citenamefont {Bauer}, \citenamefont {Chacon}, \citenamefont {Kindervater},
	\citenamefont {Muhlbauer}, \citenamefont {Pfleiderer}, \citenamefont
	{Buhrandt},\ and\ \citenamefont {et~al.}}]{Milde13}%
\BibitemOpen
\bibfield  {author} {\bibinfo {author} {\bibfnamefont {P.}~\bibnamefont
		{Milde}}, \bibinfo {author} {\bibfnamefont {D.}~\bibnamefont {Kohler}},
	\bibinfo {author} {\bibfnamefont {J.}~\bibnamefont {Seidel}}, \bibinfo
	{author} {\bibfnamefont {L.~M.}\ \bibnamefont {Eng}}, \bibinfo {author}
	{\bibfnamefont {A.}~\bibnamefont {Bauer}}, \bibinfo {author} {\bibfnamefont
		{A.}~\bibnamefont {Chacon}}, \bibinfo {author} {\bibfnamefont
		{J.}~\bibnamefont {Kindervater}}, \bibinfo {author} {\bibfnamefont
		{S.}~\bibnamefont {Muhlbauer}}, \bibinfo {author} {\bibfnamefont
		{C.}~\bibnamefont {Pfleiderer}}, \bibinfo {author} {\bibfnamefont
		{S.}~\bibnamefont {Buhrandt}}, \ and\ \bibinfo {author} {\bibnamefont
		{et~al.}},\ }\bibfield  {title} {\enquote {\bibinfo {title} {Unwinding of a
			skyrmion lattice by magnetic monopoles},}\ }\href {\doibase
	10.1126/science.1234657} {\bibfield  {journal} {\bibinfo  {journal}
		{Science}\ }\textbf {\bibinfo {volume} {340}},\ \bibinfo {pages} {1076--1080}
	(\bibinfo {year} {2013})}\BibitemShut {NoStop}%
\bibitem [{\citenamefont {Lin}\ and\ \citenamefont {Saxena}(2016)}]{Lin16}%
\BibitemOpen
\bibfield  {author} {\bibinfo {author} {\bibfnamefont {Shi-Zeng}\
		\bibnamefont {Lin}}\ and\ \bibinfo {author} {\bibfnamefont {Avadh}\
		\bibnamefont {Saxena}},\ }\bibfield  {title} {\enquote {\bibinfo {title}
		{Dynamics of {D}irac strings and monopolelike excitations in chiral magnets
			under a current drive},}\ }\href {\doibase 10.1103/physrevb.93.060401}
{\bibfield  {journal} {\bibinfo  {journal} {Phys. Rev. B}\ }\textbf {\bibinfo
		{volume} {93}},\ \bibinfo {pages} {060401} (\bibinfo {year}
	{2016})}\BibitemShut {NoStop}%
\bibitem [{\citenamefont {Pylypovskyi}\ \emph {et~al.}(2012)\citenamefont
	{Pylypovskyi}, \citenamefont {Sheka},\ and\ \citenamefont
	{Gaididei}}]{Pylypovskyi12}%
\BibitemOpen
\bibfield  {author} {\bibinfo {author} {\bibfnamefont {Oleksandr~V.}\
		\bibnamefont {Pylypovskyi}}, \bibinfo {author} {\bibfnamefont {Denis~D.}\
		\bibnamefont {Sheka}}, \ and\ \bibinfo {author} {\bibfnamefont {Yuri}\
		\bibnamefont {Gaididei}},\ }\bibfield  {title} {\enquote {\bibinfo {title}
		{Bloch point structure in a magnetic nanosphere},}\ }\href {\doibase
	10.1103/PhysRevB.85.224401} {\bibfield  {journal} {\bibinfo  {journal} {Phys.
			Rev. B}\ }\textbf {\bibinfo {volume} {85}},\ \bibinfo {pages} {224401}
	(\bibinfo {year} {2012})}\BibitemShut {NoStop}%
\bibitem [{\citenamefont {Schulz}\ \emph {et~al.}(2012)\citenamefont {Schulz},
	\citenamefont {Ritz}, \citenamefont {Bauer}, \citenamefont {Halder},
	\citenamefont {Wagner}, \citenamefont {Franz}, \citenamefont {Pfleiderer},
	\citenamefont {Everschor}, \citenamefont {Garst},\ and\ \citenamefont
	{Rosch}}]{Schulz12}%
\BibitemOpen
\bibfield  {author} {\bibinfo {author} {\bibfnamefont {T.}~\bibnamefont
		{Schulz}}, \bibinfo {author} {\bibfnamefont {R.}~\bibnamefont {Ritz}},
	\bibinfo {author} {\bibfnamefont {A.}~\bibnamefont {Bauer}}, \bibinfo
	{author} {\bibfnamefont {M.}~\bibnamefont {Halder}}, \bibinfo {author}
	{\bibfnamefont {M.}~\bibnamefont {Wagner}}, \bibinfo {author} {\bibfnamefont
		{C.}~\bibnamefont {Franz}}, \bibinfo {author} {\bibfnamefont
		{C.}~\bibnamefont {Pfleiderer}}, \bibinfo {author} {\bibfnamefont
		{K.}~\bibnamefont {Everschor}}, \bibinfo {author} {\bibfnamefont
		{M.}~\bibnamefont {Garst}}, \ and\ \bibinfo {author} {\bibfnamefont
		{A.}~\bibnamefont {Rosch}},\ }\bibfield  {title} {\enquote {\bibinfo {title}
		{Emergent electrodynamics of skyrmions in a chiral magnet},}\ }\href
{\doibase 10.1038/nphys2231} {\bibfield  {journal} {\bibinfo  {journal} {Nat
			Phys}\ }\textbf {\bibinfo {volume} {8}},\ \bibinfo {pages} {301--304}
	(\bibinfo {year} {2012})}\BibitemShut {NoStop}%
\bibitem [{\citenamefont {Neubauer}\ \emph {et~al.}(2009)\citenamefont
	{Neubauer}, \citenamefont {Pfleiderer}, \citenamefont {Binz}, \citenamefont
	{Rosch}, \citenamefont {Ritz}, \citenamefont {Niklowitz},\ and\ \citenamefont
	{Böni}}]{Neubauer09}%
\BibitemOpen
\bibfield  {author} {\bibinfo {author} {\bibfnamefont {A.}~\bibnamefont
		{Neubauer}}, \bibinfo {author} {\bibfnamefont {C.}~\bibnamefont
		{Pfleiderer}}, \bibinfo {author} {\bibfnamefont {B.}~\bibnamefont {Binz}},
	\bibinfo {author} {\bibfnamefont {A.}~\bibnamefont {Rosch}}, \bibinfo
	{author} {\bibfnamefont {R.}~\bibnamefont {Ritz}}, \bibinfo {author}
	{\bibfnamefont {P.~G.}\ \bibnamefont {Niklowitz}}, \ and\ \bibinfo {author}
	{\bibfnamefont {P.}~\bibnamefont {Böni}},\ }\bibfield  {title} {\enquote
	{\bibinfo {title} {Topological hall effect in the a phase of {MnSi}},}\
}\href {\doibase 10.1103/physrevlett.102.186602} {\bibfield  {journal}
{\bibinfo  {journal} {Phys. Rev. Lett.}\ }\textbf {\bibinfo {volume} {102}},\
\bibinfo {pages} {186602} (\bibinfo {year} {2009})}\BibitemShut {NoStop}%
\bibitem [{\citenamefont {Li}\ \emph {et~al.}(2013)\citenamefont {Li},
	\citenamefont {Kanazawa}, \citenamefont {Yu}, \citenamefont {Tsukazaki},
	\citenamefont {Kawasaki}, \citenamefont {Ichikawa}, \citenamefont {Jin},
	\citenamefont {Kagawa},\ and\ \citenamefont {Tokura}}]{Li13a}%
\BibitemOpen
\bibfield  {author} {\bibinfo {author} {\bibfnamefont {Yufan}\ \bibnamefont
		{Li}}, \bibinfo {author} {\bibfnamefont {N.}~\bibnamefont {Kanazawa}},
	\bibinfo {author} {\bibfnamefont {X.~Z.}\ \bibnamefont {Yu}}, \bibinfo
	{author} {\bibfnamefont {A.}~\bibnamefont {Tsukazaki}}, \bibinfo {author}
	{\bibfnamefont {M.}~\bibnamefont {Kawasaki}}, \bibinfo {author}
	{\bibfnamefont {M.}~\bibnamefont {Ichikawa}}, \bibinfo {author}
	{\bibfnamefont {X.~F.}\ \bibnamefont {Jin}}, \bibinfo {author} {\bibfnamefont
		{F.}~\bibnamefont {Kagawa}}, \ and\ \bibinfo {author} {\bibfnamefont
		{Y.}~\bibnamefont {Tokura}},\ }\bibfield  {title} {\enquote {\bibinfo {title}
		{Robust formation of skyrmions and topological hall effect anomaly in
			epitaxial thin films of {MnSi}},}\ }\href {\doibase
	10.1103/physrevlett.110.117202} {\bibfield  {journal} {\bibinfo  {journal}
		{Phys. Rev. Lett.}\ }\textbf {\bibinfo {volume} {110}},\ \bibinfo {pages}
	{117202} (\bibinfo {year} {2013})}\BibitemShut {NoStop}%
\bibitem [{\citenamefont {Kanazawa}\ \emph {et~al.}(2011)\citenamefont
	{Kanazawa}, \citenamefont {Onose}, \citenamefont {Arima}, \citenamefont
	{Okuyama}, \citenamefont {Ohoyama}, \citenamefont {Wakimoto}, \citenamefont
	{Kakurai}, \citenamefont {Ishiwata},\ and\ \citenamefont
	{Tokura}}]{Kanazawa11}%
\BibitemOpen
\bibfield  {author} {\bibinfo {author} {\bibfnamefont {N.}~\bibnamefont
		{Kanazawa}}, \bibinfo {author} {\bibfnamefont {Y.}~\bibnamefont {Onose}},
	\bibinfo {author} {\bibfnamefont {T.}~\bibnamefont {Arima}}, \bibinfo
	{author} {\bibfnamefont {D.}~\bibnamefont {Okuyama}}, \bibinfo {author}
	{\bibfnamefont {K.}~\bibnamefont {Ohoyama}}, \bibinfo {author} {\bibfnamefont
		{S.}~\bibnamefont {Wakimoto}}, \bibinfo {author} {\bibfnamefont
		{K.}~\bibnamefont {Kakurai}}, \bibinfo {author} {\bibfnamefont
		{S.}~\bibnamefont {Ishiwata}}, \ and\ \bibinfo {author} {\bibfnamefont
		{Y.}~\bibnamefont {Tokura}},\ }\bibfield  {title} {\enquote {\bibinfo {title}
		{Large topological hall effect in a short-period helimagnet {MnGe}},}\ }\href
{\doibase 10.1103/physrevlett.106.156603} {\bibfield  {journal} {\bibinfo
		{journal} {Phys. Rev. Lett.}\ }\textbf {\bibinfo {volume} {106}} (\bibinfo
	{year} {2011})}\BibitemShut {NoStop}%
\bibitem [{\citenamefont {Carvalho-Santos}\ and\ \citenamefont
	{Dandoloff}(2012)}]{Carvalho-Santos12}%
\BibitemOpen
\bibfield  {author} {\bibinfo {author} {\bibfnamefont {Vagson~L.}\
		\bibnamefont {Carvalho-Santos}}\ and\ \bibinfo {author} {\bibfnamefont
		{Rossen}\ \bibnamefont {Dandoloff}},\ }\bibfield  {title} {\enquote {\bibinfo
		{title} {Coupling between magnetic field and curvature in Heisenberg spins on
			surfaces with rotational symmetry},}\ }\href {\doibase
	10.1016/j.physleta.2012.10.027} {\bibfield  {journal} {\bibinfo  {journal}
		{Physics Letters A}\ }\textbf {\bibinfo {volume} {376}},\ \bibinfo {pages}
	{3551 -- 3554} (\bibinfo {year} {2012})}\BibitemShut {NoStop}%
\bibitem [{\citenamefont {Priscila S.C. Vilas-Boas}(2015)}]{priscila15}%
\BibitemOpen
\bibfield  {author} {\bibinfo {author} {\bibfnamefont {Dora Altbir Jakson M.
			Fonseca Vagson L. Carvalho-Santos}\ \bibnamefont {Priscila S.C. Vilas-Boas},
		\bibfnamefont {Ricardo G.~Elias}},\ }\bibfield  {title} {\enquote {\bibinfo
		{title} {Topological magnetic solitons on a paraboloidal shell},}\ }\href
{http://www.sciencedirect.com/science/article/pii/S0375960114010573}
{\bibfield  {journal} {\bibinfo  {journal} {Physics Letters A}\ ,\ \bibinfo
		{pages} {47–53}} (\bibinfo {year} {2015})}\BibitemShut {NoStop}%
\bibitem [{\citenamefont {Carvalho-Santos}\ \emph {et~al.}(2015)\citenamefont
	{Carvalho-Santos}, \citenamefont {Elias}, \citenamefont {Altbir},\ and\
	\citenamefont {Fonseca}}]{Carvalho-Santos15a}%
\BibitemOpen
\bibfield  {author} {\bibinfo {author} {\bibfnamefont {V.L.}\ \bibnamefont
		{Carvalho-Santos}}, \bibinfo {author} {\bibfnamefont {R.G.}\ \bibnamefont
		{Elias}}, \bibinfo {author} {\bibfnamefont {D.}~\bibnamefont {Altbir}}, \
	and\ \bibinfo {author} {\bibfnamefont {J.M.}\ \bibnamefont {Fonseca}},\
}\bibfield  {title} {\enquote {\bibinfo {title} {Stability of skyrmions on
		curved surfaces in the presence of a magnetic field},}\ }\href {\doibase
10.1016/j.jmmm.2015.04.078} {\bibfield  {journal} {\bibinfo  {journal}
	{Journal of Magnetism and Magnetic Materials}\ }\textbf {\bibinfo {volume}
	{391}},\ \bibinfo {pages} {179--183} (\bibinfo {year} {2015})}\BibitemShut
{NoStop}%
\bibitem [{\citenamefont {Carvalho-Santos}\ \emph {et~al.}(2008)\citenamefont
	{Carvalho-Santos}, \citenamefont {Moura}, \citenamefont {Moura-Melo},\ and\
	\citenamefont {Pereira}}]{Carvalho-Santos08}%
\BibitemOpen
\bibfield  {author} {\bibinfo {author} {\bibfnamefont {V.~L.}\ \bibnamefont
		{Carvalho-Santos}}, \bibinfo {author} {\bibfnamefont {A.~R.}\ \bibnamefont
		{Moura}}, \bibinfo {author} {\bibfnamefont {W.~A.}\ \bibnamefont
		{Moura-Melo}}, \ and\ \bibinfo {author} {\bibfnamefont {A.~R.}\ \bibnamefont
		{Pereira}},\ }\bibfield  {title} {\enquote {\bibinfo {title} {Topological
			spin excitations on a rigid torus},}\ }\href {\doibase
	10.1103/PhysRevB.77.134450} {\bibfield  {journal} {\bibinfo  {journal} {Phys.
			Rev. B}\ }\textbf {\bibinfo {volume} {77}},\ \bibinfo {eid} {134450}
	(\bibinfo {year} {2008})}\BibitemShut {NoStop}%
\bibitem [{\citenamefont {Carvalho-Santos}\ \emph {et~al.}(2013)\citenamefont
	{Carvalho-Santos}, \citenamefont {Apolonio},\ and\ \citenamefont
	{Oliveira-Neto}}]{Carvalho-Santos13}%
\BibitemOpen
\bibfield  {author} {\bibinfo {author} {\bibfnamefont {V.L.}\ \bibnamefont
		{Carvalho-Santos}}, \bibinfo {author} {\bibfnamefont {F.A.}\ \bibnamefont
		{Apolonio}}, \ and\ \bibinfo {author} {\bibfnamefont {N.M.}\ \bibnamefont
		{Oliveira-Neto}},\ }\bibfield  {title} {\enquote {\bibinfo {title} {On
			geometry--dependent vortex stability and topological spin excitations on
			curved surfaces with cylindrical symmetry},}\ }\href {\doibase
	10.1016/j.physleta.2013.03.028} {\bibfield  {journal} {\bibinfo  {journal}
		{Physics Letters A}\ }\textbf {\bibinfo {volume} {377}},\ \bibinfo {pages}
	{1308--1316} (\bibinfo {year} {2013})}\BibitemShut {NoStop}%
\bibitem [{\citenamefont {Cr{\'e}pieux}\ and\ \citenamefont
	{Lacroix}(1998)}]{Crepieux98}%
\BibitemOpen
\bibfield  {author} {\bibinfo {author} {\bibfnamefont {A.}~\bibnamefont
		{Cr{\'e}pieux}}\ and\ \bibinfo {author} {\bibfnamefont {C.}~\bibnamefont
		{Lacroix}},\ }\bibfield  {title} {\enquote {\bibinfo {title}
		{{Dzyaloshinsky--Moriya interactions induced by symmetry breaking at a
				surface}},}\ }\href {\doibase 10.1016/S0304-8853(97)01044-5} {\bibfield
	{journal} {\bibinfo  {journal} {Journal of Magnetism and Magnetic Materials}\
	}\textbf {\bibinfo {volume} {182}},\ \bibinfo {pages} {341--349} (\bibinfo
	{year} {1998})}\BibitemShut {NoStop}%
\bibitem [{\citenamefont {Thiaville}\ \emph {et~al.}(2012)\citenamefont
	{Thiaville}, \citenamefont {Rohart}, \citenamefont {Ju{\'e}}, \citenamefont
	{Cros},\ and\ \citenamefont {Fert}}]{Thiaville12}%
\BibitemOpen
\bibfield  {author} {\bibinfo {author} {\bibfnamefont {Andr{\'e}}\
		\bibnamefont {Thiaville}}, \bibinfo {author} {\bibfnamefont {Stanislas}\
		\bibnamefont {Rohart}}, \bibinfo {author} {\bibfnamefont {{\'E}milie}\
		\bibnamefont {Ju{\'e}}}, \bibinfo {author} {\bibfnamefont {Vincent}\
		\bibnamefont {Cros}}, \ and\ \bibinfo {author} {\bibfnamefont {Albert}\
		\bibnamefont {Fert}},\ }\bibfield  {title} {\enquote {\bibinfo {title}
		{Dynamics of {D}zyaloshinskii domain walls in ultrathin magnetic films},}\
}\href {\doibase 10.1209/0295-5075/100/57002} {\bibfield  {journal} {\bibinfo
	{journal} {EPL (Europhysics Letters)}\ }\textbf {\bibinfo {volume} {100}},\
\bibinfo {pages} {57002} (\bibinfo {year} {2012})}\BibitemShut {NoStop}%
\bibitem [{\citenamefont {Yang}\ \emph {et~al.}(2015)\citenamefont {Yang},
	\citenamefont {Thiaville}, \citenamefont {Rohart}, \citenamefont {Fert},\
	and\ \citenamefont {Chshiev}}]{Yang15}%
\BibitemOpen
\bibfield  {author} {\bibinfo {author} {\bibfnamefont {Hongxin}\ \bibnamefont
		{Yang}}, \bibinfo {author} {\bibfnamefont {Andr\'e}\ \bibnamefont
		{Thiaville}}, \bibinfo {author} {\bibfnamefont {Stanislas}\ \bibnamefont
		{Rohart}}, \bibinfo {author} {\bibfnamefont {Albert}\ \bibnamefont {Fert}}, \
	and\ \bibinfo {author} {\bibfnamefont {Mairbek}\ \bibnamefont {Chshiev}},\
}\bibfield  {title} {\enquote {\bibinfo {title} {Anatomy of
		{D}zyaloshinskii-{M}oriya interaction at $\mathrm{Co}/\mathrm{Pt}$
		interfaces},}\ }\href {\doibase 10.1103/PhysRevLett.115.267210} {\bibfield
{journal} {\bibinfo  {journal} {Phys. Rev. Lett.}\ }\textbf {\bibinfo
	{volume} {115}},\ \bibinfo {pages} {267210} (\bibinfo {year}
{2015})}\BibitemShut {NoStop}%
\bibitem [{\citenamefont {Gioia}\ and\ \citenamefont {James}(1997)}]{Gioia97}%
\BibitemOpen
\bibfield  {author} {\bibinfo {author} {\bibfnamefont {G.}~\bibnamefont
		{Gioia}}\ and\ \bibinfo {author} {\bibfnamefont {R.~D.}\ \bibnamefont
		{James}},\ }\bibfield  {title} {\enquote {\bibinfo {title} {Micromagnetics of
			very thin films},}\ }\href
{http://www.journals.royalsoc.ac.uk/openurl.asp?genre=article&id=doi:10.1098/rspa.1997.0013}
{\bibfield  {journal} {\bibinfo  {journal} {Proc. R. Soc. Lond. A}\ }\textbf
	{\bibinfo {volume} {453}},\ \bibinfo {pages} {213--223} (\bibinfo {year}
	{1997})}\BibitemShut {NoStop}%
\bibitem [{\citenamefont {Kohn}\ and\ \citenamefont
	{Slastikov}(2005{\natexlab{a}})}]{Kohn05}%
\BibitemOpen
\bibfield  {author} {\bibinfo {author} {\bibfnamefont {R.}~\bibnamefont
		{Kohn}}\ and\ \bibinfo {author} {\bibfnamefont {V.}~\bibnamefont
		{Slastikov}},\ }\bibfield  {title} {\enquote {\bibinfo {title} {Effective
			dynamics for ferromagnetic thin films: a rigorous justification},}\ }\href
{http://dx.doi.org/10.1098/rspa.2004.1342} {\bibfield  {journal} {\bibinfo
		{journal} {Proc. R. Soc. A}\ }\textbf {\bibinfo {volume} {461}},\ \bibinfo
	{pages} {143--154} (\bibinfo {year} {2005}{\natexlab{a}})}\BibitemShut
{NoStop}%
\bibitem [{\citenamefont {Kohn}\ and\ \citenamefont
	{Slastikov}(2005{\natexlab{b}})}]{Kohn05a}%
\BibitemOpen
\bibfield  {author} {\bibinfo {author} {\bibfnamefont {Robert~V.}\
		\bibnamefont {Kohn}}\ and\ \bibinfo {author} {\bibfnamefont {Valeriy~V.}\
		\bibnamefont {Slastikov}},\ }\bibfield  {title} {\enquote {\bibinfo {title}
		{Another thin-film limit of micromagnetics},}\ }\href
{http://dx.doi.org/10.1007/s00205-005-0372-7} {\bibfield  {journal} {\bibinfo
		{journal} {Archive for Rational Mechanics and Analysis}\ }\textbf {\bibinfo
		{volume} {178}},\ \bibinfo {pages} {227--245} (\bibinfo {year}
	{2005}{\natexlab{b}})}\BibitemShut {NoStop}%
\bibitem [{\citenamefont {Slastikov}(2005)}]{Slastikov05}%
\BibitemOpen
\bibfield  {author} {\bibinfo {author} {\bibfnamefont {Valery}\ \bibnamefont
		{Slastikov}},\ }\bibfield  {title} {\enquote {\bibinfo {title}
		{Micromagnetism of thin shells},}\ }\href {\doibase
	10.1142/S021820250500087X} {\bibfield  {journal} {\bibinfo  {journal}
		{Mathematical Models and Methods in Applied Sciences}\ }\textbf {\bibinfo
		{volume} {15}},\ \bibinfo {pages} {1469--1487} (\bibinfo {year} {2005})}
\BibitemShut {NoStop}%
\bibitem [{Note3()}]{Note3}%
\BibitemOpen
\bibinfo {note} {Before applying the limit $R\to \infty $ one should make a
	change of independent variable $\vartheta =\rho /R$, where $\rho $ is a
	distance along a meridian direction.}\BibitemShut {Stop}%
\bibitem [{\citenamefont {Kiselev}\ \emph {et~al.}(2011)\citenamefont
	{Kiselev}, \citenamefont {Bogdanov}, \citenamefont {Sch{\"a}fer},\ and\
	\citenamefont {R{\"o}{\ss}ler}}]{Kiselev11}%
\BibitemOpen
\bibfield  {author} {\bibinfo {author} {\bibfnamefont {N.~S.}\ \bibnamefont
		{Kiselev}}, \bibinfo {author} {\bibfnamefont {A.~N.}\ \bibnamefont
		{Bogdanov}}, \bibinfo {author} {\bibfnamefont {R.}~\bibnamefont
		{Sch{\"a}fer}}, \ and\ \bibinfo {author} {\bibfnamefont {U.~K.}\ \bibnamefont
		{R{\"o}{\ss}ler}},\ }\bibfield  {title} {\enquote {\bibinfo {title} {Chiral
			skyrmions in thin magnetic films: new objects for magnetic storage
			technologies?}}\ }\href {http://stacks.iop.org/0022-3727/44/i=39/a=392001}
{\bibfield  {journal} {\bibinfo  {journal} {Journal of Physics D: Applied
			Physics}\ }\textbf {\bibinfo {volume} {44}},\ \bibinfo {pages} {392001}
	(\bibinfo {year} {2011})}\BibitemShut {NoStop}%
\bibitem [{\citenamefont {Streubel}\ \emph {et~al.}(2015)\citenamefont
	{Streubel}, \citenamefont {Han}, \citenamefont {Im}, \citenamefont {Kronast},
	\citenamefont {R{\"o\ss}ler}, \citenamefont {Radu}, \citenamefont {Abrudan},
	\citenamefont {Lin}, \citenamefont {Schmidt}, \citenamefont {Fischer},\ and\
	\citenamefont {Makarov}}]{Streubel15}%
\BibitemOpen
\bibfield  {author} {\bibinfo {author} {\bibfnamefont {Robert}\ \bibnamefont
		{Streubel}}, \bibinfo {author} {\bibfnamefont {Luyang}\ \bibnamefont {Han}},
	\bibinfo {author} {\bibfnamefont {Mi-Young}\ \bibnamefont {Im}}, \bibinfo
	{author} {\bibfnamefont {Florian}\ \bibnamefont {Kronast}}, \bibinfo {author}
	{\bibfnamefont {Ulrich~K.}\ \bibnamefont {R{\"o\ss}ler}}, \bibinfo {author}
	{\bibfnamefont {Florin}\ \bibnamefont {Radu}}, \bibinfo {author}
	{\bibfnamefont {Radu}\ \bibnamefont {Abrudan}}, \bibinfo {author}
	{\bibfnamefont {Gungun}\ \bibnamefont {Lin}}, \bibinfo {author}
	{\bibfnamefont {Oliver~G.}\ \bibnamefont {Schmidt}}, \bibinfo {author}
	{\bibfnamefont {Peter}\ \bibnamefont {Fischer}}, \ and\ \bibinfo {author}
	{\bibfnamefont {Denys}\ \bibnamefont {Makarov}},\ }\bibfield  {title}
{\enquote {\bibinfo {title} {Manipulating topological states by imprinting
			non-collinear spin textures},}\ }\href {\doibase 10.1038/srep08787}
{\bibfield  {journal} {\bibinfo  {journal} {Scientific Reports}\ }\textbf
	{\bibinfo {volume} {5}},\ \bibinfo {pages} {8787} (\bibinfo {year}
	{2015})}\BibitemShut {NoStop}%
\bibitem [{\citenamefont {Beg}\ \emph {et~al.}(2015)\citenamefont {Beg},
	\citenamefont {Carey}, \citenamefont {Wang}, \citenamefont
	{Cort{\'{e}}s-Ortu{\~{n}}o}, \citenamefont {Vousden}, \citenamefont
	{Bisotti}, \citenamefont {Albert}, \citenamefont {Chernyshenko},
	\citenamefont {Hovorka}, \citenamefont {Stamps},\ and\ \citenamefont
	{Fangohr}}]{Beg15}%
\BibitemOpen
\bibfield  {author} {\bibinfo {author} {\bibfnamefont {Marijan}\ \bibnamefont
		{Beg}}, \bibinfo {author} {\bibfnamefont {Rebecca}\ \bibnamefont {Carey}},
	\bibinfo {author} {\bibfnamefont {Weiwei}\ \bibnamefont {Wang}}, \bibinfo
	{author} {\bibfnamefont {David}\ \bibnamefont {Cort{\'{e}}s-Ortu{\~{n}}o}},
	\bibinfo {author} {\bibfnamefont {Mark}\ \bibnamefont {Vousden}}, \bibinfo
	{author} {\bibfnamefont {Marc-Antonio}\ \bibnamefont {Bisotti}}, \bibinfo
	{author} {\bibfnamefont {Maximilian}\ \bibnamefont {Albert}}, \bibinfo
	{author} {\bibfnamefont {Dmitri}\ \bibnamefont {Chernyshenko}}, \bibinfo
	{author} {\bibfnamefont {Ondrej}\ \bibnamefont {Hovorka}}, \bibinfo {author}
	{\bibfnamefont {Robert~L.}\ \bibnamefont {Stamps}}, \ and\ \bibinfo {author}
	{\bibfnamefont {Hans}\ \bibnamefont {Fangohr}},\ }\bibfield  {title}
{\enquote {\bibinfo {title} {Ground state search, hysteretic behaviour, and
			reversal mechanism of skyrmionic textures in confined helimagnetic
			nanostructures},}\ }\href {\doibase 10.1038/srep17137} {\bibfield  {journal}
	{\bibinfo  {journal} {Scientific Reports}\ }\textbf {\bibinfo {volume} {5}},\
	\bibinfo {pages} {17137} (\bibinfo {year} {2015})}\BibitemShut {NoStop}%
\bibitem [{\citenamefont {Bogdanov}\ and\ \citenamefont
	{Hubert}(1994{\natexlab{b}})}]{Bogdanov94a}%
\BibitemOpen
\bibfield  {author} {\bibinfo {author} {\bibfnamefont {A.}~\bibnamefont
		{Bogdanov}}\ and\ \bibinfo {author} {\bibfnamefont {A.}~\bibnamefont
		{Hubert}},\ }\bibfield  {title} {\enquote {\bibinfo {title} {The properties
			of isolated magnetic vortices},}\ }\href {\doibase 10.1002/pssb.2221860223}
{\bibfield  {journal} {\bibinfo  {journal} {Physica status solidi (b)}\
	}\textbf {\bibinfo {volume} {186}},\ \bibinfo {pages} {527--543} (\bibinfo
	{year} {1994}{\natexlab{b}})}\BibitemShut {NoStop}%
\bibitem [{\citenamefont {Tretiakov}\ and\ \citenamefont
	{Tchernyshyov}(2007)}]{Tretiakov07}%
\BibitemOpen
\bibfield  {author} {\bibinfo {author} {\bibfnamefont {O.~A.}\ \bibnamefont
		{Tretiakov}}\ and\ \bibinfo {author} {\bibfnamefont {O.}~\bibnamefont
		{Tchernyshyov}},\ }\bibfield  {title} {\enquote {\bibinfo {title} {Vortices
			in thin ferromagnetic films and the skyrmion number},}\ }\href
{http://link.aps.org/abstract/PRB/v75/e012408} {\bibfield  {journal}
	{\bibinfo  {journal} {Phys. Rev. B}\ }\textbf {\bibinfo {volume} {75}},\
	\bibinfo {eid} {012408} (\bibinfo {year} {2007})}\BibitemShut {NoStop}%
\bibitem [{\citenamefont {Hertel}\ and\ \citenamefont
	{Schneider}(2006)}]{Hertel06}%
\BibitemOpen
\bibfield  {author} {\bibinfo {author} {\bibfnamefont {Riccardo}\
		\bibnamefont {Hertel}}\ and\ \bibinfo {author} {\bibfnamefont {Claus~M.}\
		\bibnamefont {Schneider}},\ }\bibfield  {title} {\enquote {\bibinfo {title}
		{Exchange explosions: Magnetization dynamics during vortex-antivortex
			annihilation},}\ }\href {http://link.aps.org/abstract/PRL/v97/e177202}
{\bibfield  {journal} {\bibinfo  {journal} {Phys. Rev. Lett.}\ }\textbf
	{\bibinfo {volume} {97}},\ \bibinfo {eid} {177202} (\bibinfo {year}
	{2006})}\BibitemShut {NoStop}%
\bibitem [{\citenamefont {Streubel}\ \emph {et~al.}(2016)\citenamefont
	{Streubel}, \citenamefont {Fischer}, \citenamefont {Kronast}, \citenamefont
	{Kravchuk}, \citenamefont {Sheka}, \citenamefont {Gaididei}, \citenamefont
	{Schmidt},\ and\ \citenamefont {Makarov}}]{Streubel16a}%
\BibitemOpen
\bibfield  {author} {\bibinfo {author} {\bibfnamefont {Robert}\ \bibnamefont
		{Streubel}}, \bibinfo {author} {\bibfnamefont {Peter}\ \bibnamefont
		{Fischer}}, \bibinfo {author} {\bibfnamefont {Florian}\ \bibnamefont
		{Kronast}}, \bibinfo {author} {\bibfnamefont {Volodymyr~P.}\ \bibnamefont
		{Kravchuk}}, \bibinfo {author} {\bibfnamefont {Denis~D.}\ \bibnamefont
		{Sheka}}, \bibinfo {author} {\bibfnamefont {Yuri}\ \bibnamefont {Gaididei}},
	\bibinfo {author} {\bibfnamefont {Oliver~G.}\ \bibnamefont {Schmidt}}, \ and\
	\bibinfo {author} {\bibfnamefont {Denys}\ \bibnamefont {Makarov}},\
}\bibfield  {title} {\enquote {\bibinfo {title} {Magnetism in curved
		geometries (topical review)},}\ }\href {\doibase
10.1088/0022-3727/49/36/363001} {\bibfield  {journal} {\bibinfo  {journal}
	{Journal of Physics D: Applied Physics}\ }\textbf {\bibinfo {volume} {49}},\
\bibinfo {pages} {363001} (\bibinfo {year} {2016})}\BibitemShut {NoStop}%
\bibitem [{\citenamefont {Zhang}\ \emph {et~al.}(2009)\citenamefont {Zhang},
	\citenamefont {Ding}, \citenamefont {Chow}, \citenamefont {Ran},\ and\
	\citenamefont {Yi}}]{Zhang09a}%
\BibitemOpen
\bibfield  {author} {\bibinfo {author} {\bibfnamefont {HaiTao}\ \bibnamefont
		{Zhang}}, \bibinfo {author} {\bibfnamefont {Jun}\ \bibnamefont {Ding}},
	\bibinfo {author} {\bibfnamefont {GanMoog}\ \bibnamefont {Chow}}, \bibinfo
	{author} {\bibfnamefont {Min}\ \bibnamefont {Ran}}, \ and\ \bibinfo {author}
	{\bibfnamefont {JiaBao}\ \bibnamefont {Yi}},\ }\bibfield  {title} {\enquote
	{\bibinfo {title} {Engineering magnetic properties of Ni nanoparticles by
			non-magnetic cores},}\ }\href {\doibase 10.1021/cm902114d} {\bibfield
	{journal} {\bibinfo  {journal} {Chemistry of Materials}\ }\textbf {\bibinfo
		{volume} {21}},\ \bibinfo {pages} {5222--5228} (\bibinfo {year}
	{2009})}\BibitemShut {NoStop}%
\bibitem [{\citenamefont {Cabot}\ \emph {et~al.}(2009)\citenamefont {Cabot},
	\citenamefont {Alivisatos}, \citenamefont {Puntes}, \citenamefont {Balcells},
	\citenamefont {Iglesias},\ and\ \citenamefont {Labarta}}]{Cabot09}%
\BibitemOpen
\bibfield  {author} {\bibinfo {author} {\bibfnamefont {Andreu}\ \bibnamefont
		{Cabot}}, \bibinfo {author} {\bibfnamefont {A.~Paul}\ \bibnamefont
		{Alivisatos}}, \bibinfo {author} {\bibfnamefont {V{\'{\i}}ctor~F.}\
		\bibnamefont {Puntes}}, \bibinfo {author} {\bibfnamefont {Llu{\'{\i}}s}\
		\bibnamefont {Balcells}}, \bibinfo {author} {\bibfnamefont {{\`{O}}scar}\
		\bibnamefont {Iglesias}}, \ and\ \bibinfo {author} {\bibfnamefont
		{Am{\'{\i}}lcar}\ \bibnamefont {Labarta}},\ }\bibfield  {title} {\enquote
	{\bibinfo {title} {Magnetic domains and surface effects in hollow maghemite
			nanoparticles},}\ }\href {\doibase 10.1103/physrevb.79.094419} {\bibfield
	{journal} {\bibinfo  {journal} {Phys. Rev. B}\ }\textbf {\bibinfo {volume}
		{79}} (\bibinfo {year} {2009}),\ 10.1103/physrevb.79.094419}\BibitemShut
{NoStop}%
\bibitem [{\citenamefont {Gong}\ \emph {et~al.}(2014)\citenamefont {Gong},
	\citenamefont {Kirkeminde}, \citenamefont {Skomski}, \citenamefont {Cui},\
	and\ \citenamefont {Ren}}]{Gong14}%
\BibitemOpen
\bibfield  {author} {\bibinfo {author} {\bibfnamefont {Maogang}\ \bibnamefont
		{Gong}}, \bibinfo {author} {\bibfnamefont {Alec}\ \bibnamefont {Kirkeminde}},
	\bibinfo {author} {\bibfnamefont {Ralph}\ \bibnamefont {Skomski}}, \bibinfo
	{author} {\bibfnamefont {Jun}\ \bibnamefont {Cui}}, \ and\ \bibinfo {author}
	{\bibfnamefont {Shenqiang}\ \bibnamefont {Ren}},\ }\bibfield  {title}
{\enquote {\bibinfo {title} {Template-directed FeCo nanoshells on AuCu},}\
}\href {\doibase 10.1002/smll.201401049} {\bibfield  {journal} {\bibinfo
	{journal} {Small}\ }\textbf {\bibinfo {volume} {10}},\ \bibinfo {pages}
{4118--4122} (\bibinfo {year} {2014})}\BibitemShut {NoStop}%
\bibitem [{\citenamefont {Albrecht}\ \emph {et~al.}(2005)\citenamefont
	{Albrecht}, \citenamefont {Hu}, \citenamefont {Guhr}, \citenamefont
	{Ulbrich}, \citenamefont {Boneberg}, \citenamefont {Leiderer},\ and\
	\citenamefont {Schatz}}]{Albrecht05}%
\BibitemOpen
\bibfield  {author} {\bibinfo {author} {\bibfnamefont {Manfred}\ \bibnamefont
		{Albrecht}}, \bibinfo {author} {\bibfnamefont {Guohan}\ \bibnamefont {Hu}},
	\bibinfo {author} {\bibfnamefont {Ildico~L.}\ \bibnamefont {Guhr}}, \bibinfo
	{author} {\bibfnamefont {Till~C.}\ \bibnamefont {Ulbrich}}, \bibinfo {author}
	{\bibfnamefont {Johannes}\ \bibnamefont {Boneberg}}, \bibinfo {author}
	{\bibfnamefont {Paul}\ \bibnamefont {Leiderer}}, \ and\ \bibinfo {author}
	{\bibfnamefont {Gunter}\ \bibnamefont {Schatz}},\ }\bibfield  {title}
{\enquote {\bibinfo {title} {Magnetic multilayers on nanospheres},}\ }\href
{\doibase 10.1038/nmat1324} {\bibfield  {journal} {\bibinfo  {journal} {Nat
			Mater}\ }\textbf {\bibinfo {volume} {4}},\ \bibinfo {pages} {203--206}
	(\bibinfo {year} {2005})}\BibitemShut {NoStop}%
\bibitem [{\citenamefont {Ulbrich}\ \emph {et~al.}(2006)\citenamefont
	{Ulbrich}, \citenamefont {Makarov}, \citenamefont {Hu}, \citenamefont {Guhr},
	\citenamefont {Suess}, \citenamefont {Schrefl},\ and\ \citenamefont
	{Albrecht}}]{Ulbrich06}%
\BibitemOpen
\bibfield  {author} {\bibinfo {author} {\bibfnamefont {T.~C.}\ \bibnamefont
		{Ulbrich}}, \bibinfo {author} {\bibfnamefont {D.}~\bibnamefont {Makarov}},
	\bibinfo {author} {\bibfnamefont {G.}~\bibnamefont {Hu}}, \bibinfo {author}
	{\bibfnamefont {I.~L.}\ \bibnamefont {Guhr}}, \bibinfo {author}
	{\bibfnamefont {D.}~\bibnamefont {Suess}}, \bibinfo {author} {\bibfnamefont
		{T.}~\bibnamefont {Schrefl}}, \ and\ \bibinfo {author} {\bibfnamefont
		{M.}~\bibnamefont {Albrecht}},\ }\bibfield  {title} {\enquote {\bibinfo
		{title} {Magnetization reversal in a novel gradient nanomaterial},}\ }\href
{\doibase 10.1103/PhysRevLett.96.077202} {\bibfield  {journal} {\bibinfo
		{journal} {Phys. Rev. Lett.}\ }\textbf {\bibinfo {volume} {96}},\ \bibinfo
	{pages} {077202} (\bibinfo {year} {2006})}\BibitemShut {NoStop}%
\bibitem [{\citenamefont {Makarov}\ \emph {et~al.}(2008)\citenamefont
	{Makarov}, \citenamefont {Ure{\~{n}}a}, \citenamefont {Schmidt},
	\citenamefont {Liscio}, \citenamefont {Maret}, \citenamefont {Brombacher},
	\citenamefont {Schulze}, \citenamefont {Hietschold},\ and\ \citenamefont
	{Albrecht}}]{Makarov08}%
\BibitemOpen
\bibfield  {author} {\bibinfo {author} {\bibfnamefont {D.}~\bibnamefont
		{Makarov}}, \bibinfo {author} {\bibfnamefont {E.~Berm{\'{u}}dez}\
		\bibnamefont {Ure{\~{n}}a}}, \bibinfo {author} {\bibfnamefont {O.~G.}\
		\bibnamefont {Schmidt}}, \bibinfo {author} {\bibfnamefont {F.}~\bibnamefont
		{Liscio}}, \bibinfo {author} {\bibfnamefont {M.}~\bibnamefont {Maret}},
	\bibinfo {author} {\bibfnamefont {C.}~\bibnamefont {Brombacher}}, \bibinfo
	{author} {\bibfnamefont {S.}~\bibnamefont {Schulze}}, \bibinfo {author}
	{\bibfnamefont {M.}~\bibnamefont {Hietschold}}, \ and\ \bibinfo {author}
	{\bibfnamefont {M.}~\bibnamefont {Albrecht}},\ }\bibfield  {title} {\enquote
	{\bibinfo {title} {Nanopatterned {CoPt} alloys with perpendicular magnetic
			anisotropy},}\ }\href {\doibase 10.1063/1.2993334} {\bibfield  {journal}
	{\bibinfo  {journal} {Appl. Phys. Lett.}\ }\textbf {\bibinfo {volume} {93}},\
	\bibinfo {pages} {153112} (\bibinfo {year} {2008})}\BibitemShut {NoStop}%
\bibitem [{\citenamefont {Makarov}\ \emph {et~al.}(2009)\citenamefont
	{Makarov}, \citenamefont {Klimenta}, \citenamefont {Fischer}, \citenamefont
	{Liscio}, \citenamefont {Schulze}, \citenamefont {Hietschold}, \citenamefont
	{Maret},\ and\ \citenamefont {Albrecht}}]{Makarov09}%
\BibitemOpen
\bibfield  {author} {\bibinfo {author} {\bibfnamefont {D.}~\bibnamefont
		{Makarov}}, \bibinfo {author} {\bibfnamefont {F.}~\bibnamefont {Klimenta}},
	\bibinfo {author} {\bibfnamefont {S.}~\bibnamefont {Fischer}}, \bibinfo
	{author} {\bibfnamefont {F.}~\bibnamefont {Liscio}}, \bibinfo {author}
	{\bibfnamefont {S.}~\bibnamefont {Schulze}}, \bibinfo {author} {\bibfnamefont
		{M.}~\bibnamefont {Hietschold}}, \bibinfo {author} {\bibfnamefont
		{M.}~\bibnamefont {Maret}}, \ and\ \bibinfo {author} {\bibfnamefont
		{M.}~\bibnamefont {Albrecht}},\ }\bibfield  {title} {\enquote {\bibinfo
		{title} {Nonepitaxially grown nanopatterned co--pt alloys with out-of-plane
			magnetic anisotropy},}\ }\href {\doibase 10.1063/1.3260243} {\bibfield
	{journal} {\bibinfo  {journal} {J.~Appl. Phys.}\ }\textbf {\bibinfo {volume}
		{106}},\ \bibinfo {eid} {114322} (\bibinfo {year} {2009})}\BibitemShut
{NoStop}%
\bibitem [{\citenamefont {Dubrovin}\ \emph {et~al.}(1984)\citenamefont
	{Dubrovin}, \citenamefont {Fomenko},\ and\ \citenamefont
	{Novikov}}]{Dubrovin84p1}%
\BibitemOpen
\bibfield  {author} {\bibinfo {author} {\bibfnamefont {B.~A.}\ \bibnamefont
		{Dubrovin}}, \bibinfo {author} {\bibfnamefont {A.~T.}\ \bibnamefont
		{Fomenko}}, \ and\ \bibinfo {author} {\bibfnamefont {S.~P.}\ \bibnamefont
		{Novikov}},\ }\href@noop {} {\emph {\bibinfo {title} {Modern Geometry —
			Methods and Applications: Part I. The Geometry of Surfaces, Transformation
			Groups, and Fields}}},\ \bibinfo {edition} {2nd}\ ed.,\ \bibinfo {series}
{Graduate Texts in Mathematics 93}, Vol.~\bibinfo {volume} {1}\ (\bibinfo
{publisher} {Springer New York},\ \bibinfo {year} {1984})\BibitemShut
{NoStop}%
\bibitem [{\citenamefont {Fischbacher}\ \emph {et~al.}(2007)\citenamefont
	{Fischbacher}, \citenamefont {Franchin}, \citenamefont {Bordignon},\ and\
	\citenamefont {Fangohr}}]{Fischbacher07}%
\BibitemOpen
\bibfield  {author} {\bibinfo {author} {\bibfnamefont {Thomas}\ \bibnamefont
		{Fischbacher}}, \bibinfo {author} {\bibfnamefont {Matteo}\ \bibnamefont
		{Franchin}}, \bibinfo {author} {\bibfnamefont {Giuliano}\ \bibnamefont
		{Bordignon}}, \ and\ \bibinfo {author} {\bibfnamefont {Hans}\ \bibnamefont
		{Fangohr}},\ }\bibfield  {title} {\enquote {\bibinfo {title} {A systematic
			approach to multiphysics extensions of finite-element-based micromagnetic
			simulations: Nmag},}\ }\href {\doibase 10.1109/tmag.2007.893843} {\bibfield
	{journal} {\bibinfo  {journal} {IEEE Trans. Magn.}\ }\textbf {\bibinfo
		{volume} {43}},\ \bibinfo {pages} {2896--2898} (\bibinfo {year}
	{2007})}\BibitemShut {NoStop}%
\end{thebibliography}
%

%

\end{document}